\documentclass[12pt]{article}
\usepackage{epsfig, amsmath, amssymb}
\usepackage{graphicx,psfrag}
\usepackage{xcolor} 
\usepackage{float}
\newcommand{\nn}{\nonumber}
\newcommand{\be}{\begin{equation}}
\newcommand{\ee}{\end{equation}}
\newcommand{\ben}{\begin{equation}}
\newcommand{\een}{\end{equation}}
\newcommand{\bea}{\begin{eqnarray}}
\newcommand{\eea}{\end{eqnarray}}
\newcommand{\bA}{\begin{array}}
\newcommand{\eA}{\end{array}}
\newcommand{\bc}{\begin{center}}
\newcommand{\ec}{\end{center}}
\newcommand{\al}{\alpha}

\newcommand{\ra}{\rightarrow}
\newcommand{\del}{\partial}

\newcommand{\ie}{{\it i.e.}}
\newcommand{\eg}{{\it e.g.}}

\newcommand{\Nf}{${\cal N}{=}4$}
\newcommand{\Nt}{${\cal N}{=}2$}

\newcommand{\vn}{{\vec\nabla}}
\newcommand{\vB}{{\vec B}}
\newcommand{\vE}{{\vec E}}

\newcommand{\vx}{{\vec x}}

\newcommand{\vr}{{\vec r}}
\newcommand{\vrho}{{\vec \rho}}

\def\BR{{\mathbb R}}

\numberwithin{equation}{section}

\begin{document}


\begin{titlepage}

\bc

\hfill 
\\         [22mm]


{\Huge M5-brane prongs,\ \  string soliton bound states  \\ [2mm]
and wall-crossing} 
\vspace{16mm}

{\large  Varun Gupta,\ \ K.~Narayan} \\
\vspace{3mm}
{\small \it Chennai Mathematical Institute, \\}
{\small \it SIPCOT IT Park, Siruseri 603103, India.\\}

\ec
\vspace{30mm}

\begin{abstract}
We study abelian M5-brane field configurations representing BPS bound
states of self-dual string solitons whose locations correspond to the
endlines of M2-branes ending on the M5-branes. The BPS equations are
obtained from appropriate Bogomolny completion of the effective
abelian low energy functional with two transverse scalars, using two
vectors representing the directions along which these endline strings
extend. Then we impose boundary conditions on the scalars near the
string soliton cores. This leads to a molecule-like equilibrium
structure of two non-parallel string solitons at fixed transverse
separations, with the M5-brane ``prong'' deformations comprising two
``spikes'', each shaped like a ridge. The resulting picture becomes
increasingly accurate as one approaches the wall of marginal
stability, on which these states decay. There are various parallels
with wall-crossing phenomena for string web configurations obtained
from D3-brane deformations.
\end{abstract}

\end{titlepage}

{\tiny 
\begin{tableofcontents}
\end{tableofcontents}
}


\section{Introduction}

M5-branes are fascinating objects in M-theory. The theory on a single
M5-brane in flat space is free, comprising a self-dual anti-symmetric
3-form tensor field strength and five scalar fields and fermions. A
stack of coincident M5-branes is described by an interacting
non-abelian 3-form theory with massless scalars. Since the self-dual
3-form field strength arises from a 2-form potential which couples to
a string-like object, the M5-brane theory appears to be a theory of
interacting self-dual strings: these strings can be thought of as the
endlines of M2-branes that end on the M5-brane. A succinct description
of various aspects appears in \cite{Seiberg:1997ax} and some early
nice papers include
\eg\ \cite{Witten:1995zh,Strominger:1995ac,Townsend:1995af,Seiberg:1996vs}.
While we do not understand this theory fully, we do know several
aspects from various points of view: a classic old review is
\cite{Townsend:1997wg}. More recent reviews of several aspects include
\eg\ \cite{Bagger:2012jb,Lambert:2019khh}. The extensive
investigations of holography
\cite{Maldacena:1997re,Gubser:1998bc,Witten:1998qj} over the last
several years suggest that the M5-brane $(2,0)$ theory can be aptly
regarded as an abstract conformal field theory (see
\eg\ \cite{Beem:2015aoa} for a bootstrap formulation) and the
self-dual strings above are then solitonic objects.

When the M5-branes in a stack are separated, the M2-branes stretched
between them lead to charged massive states in the M5-theory ``Higgsed'',
loosely speaking (more precisely, the M5-theory on the tensor
branch). A single M2-brane stretched between two parallel separated
M5-branes ends on a line in the M5-worldvolume thus giving rise to a
half-BPS self-dual string soliton state in the worldvolume M5-brane
theory \cite{Howe:1997ue,Gauntlett:1997ss}. As a field configuration
in the full ``Higgsed'' M5-brane nonabelian tensor theory, this is perhaps
a line-charge analog of the BPS 't Hooft-Polyakov monopole in super
Yang-Mills theory (see \eg\ \cite{Harvey:1996ur} for a review), and
can be described at least in the low energy abelian ``Higgsed'' theory in
a simple way (as we will review below).  Consider the theory of three
parallel M5-branes: two non-parallel M2-branes stretching between one
M5 and the other two energetically lead to a nontrivial M5-M2-M2 bound
state. These sorts of states can be recognized readily in a different
regime. They simply correspond to nontrivial string webs stretching
between three parallel separated D3-branes in the IIB string theory
obtained by compactifying M-theory on $T^2$: see
\cite{Schwarz:1996bh}, where these M2-brane webs in M-theory were in
fact already anticipated. The M5-branes become D3-branes while the
M2-branes wrapping appropriate $T^2$-cycles become various
$(p,q)$-strings (as reviewed in \eg\ \cite{Polchinski:1998rr}). The
line charges from the M2-branes ending on the M5-brane compactify to
point charges in the Higgsed gauge theory on the D3-branes.

These string webs can in fact be obtained as nontrivial brane
deformations in the D3-brane worldvolume super Yang-Mills theory. As a
BPS state, a single string ending on a single D3-brane is described in
the low energy abelian theory by the BPS bound equation $\vE=\nabla X$
with $\vE$ the electric field and $X$ one of the transverse
scalars. The resulting field configuration $X\sim {q_e\over
  |\vr-\vr_0|}-X_0$ can be interpreted as a ``brane spike''
worldvolume deformation emanating from the D3-brane at the location of
the charge \cite{Callan:1997kz}.  Likewise string web states can also
be described as more general ``brane prongs'' in the low energy
abelian theory of three or more Higgsed D3-branes
\cite{Argyres:2001pv,Argyres:2000xs,Narayan:2007tx}: this low energy
description was developed from several previous investigations of such
states including \cite{Gauntlett:1999xz} as well as
\eg\ \cite{Bergman:1997yw}-\cite{Ritz:2000xa}\ (see also the review
\cite{Weinberg:2006rq}).  These are nontrivial molecule-like dyon
bound states in the Higgsed vacua of the \Nf\ $U(N)$ super Yang-Mills
theory on the D3-branes: they decay across walls of marginal stability
and the wall-crossing phenomena are in fact well-described by these
brane prong field configurations in the low energy abelian theory (the
qualitative picture has parallels with similar observations in BPS
black hole bound states \cite{Denef:2000nb}).

Motivated by these investigations, in this paper, we consider abelian
M5-brane field configurations representing BPS bound states of
self-dual string solitons whose locations correspond to the endlines
of M2-branes ending on the M5-branes. A single M2-brane ending on a
single M5-brane leads to a ${1\over 2}$-BPS state described by a
ridge-shaped \emph{``spike''} deformation of the M5-brane with one
transverse scalar excited as $X\sim {q\over
  |\vrho-\vrho_0|^2}-X_0$\,. This M5-brane ``spike'' field
configuration characterizes the self-dual string that is the endline
of the M2-brane ending on the M5-brane and was discussed in
\cite{Howe:1997ue,Gauntlett:1997ss} (reviewed in
\cite{Townsend:1997wg}).  Appropriate combinations of multiple
M2-branes give ${1\over 4}$-BPS states described by M5-brane
\emph{``prong''} deformations: the structure of these is more
intricate. They can be thought of as a superposition of two spikes
corresponding to a bound state of  two non-parallel string solitons at
fixed separation transverse to their extended directions, which
dovetails with nontrivial boundary conditions on the scalar moduli in
the vicinity of the string soliton ``cores''. Under $M\ra IIB$ duality
obtained from compactifying M-theory on $T^2$, they are expected to
reduce to D3-brane prongs representing string webs stretched between
multiple parallel non-coincident D3-branes described above.

The BPS bound equations in the M5-brane are obtained by performing a
Bogomolny completion of the M5-brane energy functional comprising a
free self-dual 3-form field strength $H_{abc}$ and two transverse
scalars. This was already set up in a preliminary manner in
\cite{Gauntlett:1999xz}: here we will develop this more elaborately.
It is convenient to recast $H_{abc}$ in terms of a spatially ``dual''
2-form object ${\tilde H}_{ab}\equiv \star_5 H$\ (where $a,b$ are
5-space indices alone, reflecting the self-duality of $H$): this is
then Bogomolny-squared with the scalar derivatives using two vectors
$\zeta^a$ to obtain BPS bound equations (in one form, these were
obtained from supersymmetry in \cite{Gauntlett:1999aw}; our analysis,
complementing that, is more fuelled by wall-crossing considerations
and the D3-brane string web descriptions). These $\zeta$-vectors have
support only in a 2-dim ($4$-$5$) subspace of the 5-dim worldvolume of
the M5-brane: this simulates the intuition that upon compactifying
this 2-subspace as $T^2$ the line charges wrap the $\zeta$-directions
becoming point charges in the noncompact 3-dimensions. In this light,
these equations can be decomposed into components along the $4-$
and/or $5-$ directions, and some of these components then resemble the
BPS equations for the D3-brane.  Another crucial ingredient in the
M5-description is the fact that the scalar field configurations are
subject to ``line source conditions'' encoding translation invariance
along the string, reflected in the ridge-shape. These alongwith the
BPS bound equations and the moduli boundary conditions lead us to the
spike superpositions in the prong field configurations.

The moduli boundary conditions in the vicinity of the string cores we
impose are Dirichlet boundary conditions on the scalars at the closest
point of approach of the two strings: these imply fixed separations
transverse to the extended directions of the strings. For generic
finite separation, these lead to nontrivial variations in the moduli
as we go far along the strings, which amounts to nontrivial M5-brane
bendings. However in the limit of approaching the wall of marginal
stability, we show that these variations become negligible and the
moduli values become near constant all along the strings. Thus our
formulation reveals self-consistent approximate configurations which
become increasingly accurate in the marginal stability limit, as we
will describe elaborately.

In sec.~\ref{sec:Bogo}, we describe the Bogomolny completion in the
M5-brane low energy abelian theory, while sec.~\ref{sec:m5m2}
describes the simplest M5-brane prong configurations and the
corresponding string soliton bound states with the effective 1-string
tension and the wall-crossing limit in sec.~\ref{sec:tension}\
(with some further comments in sec.~\ref{sec:Bog1str-os}). More
general states appear in sec.~\ref{sec:m5m2gen}. In
sec.~\ref{M5s-higgsed} we discuss embedding these field configurations
in multiple abelian ``Higgsed'' M5-brane theories. A Discussion
comprising an overview and related questions appears in \ref{sec:Disc},
and some technical details in Appendices.

\section{Bogomolny completion in M5 effective field theory}\label{sec:Bogo}

To describe the M5-brane spatial worldvolume for the field
configurations and states to follow, we will find it useful to define
the notation
\bea\label{notationrhozeta}
\vrho =(x_1,x_2,x_3,x_4,x_5) \equiv (\vr, \vx)\ ; &&
\rho_X\equiv (\vr,x_5)\ ; \qquad \rho_Y\equiv (\vr,x_4)\ ,\nn\\
\rho_\zeta \equiv (\vr,\vx) , &{\rm with}& \vx\cdot\zeta=0\ .
\eea
\ie\ $\vrho$ is the coordinate describing the 5-dimensional spatial
worldvolume of the M5-brane which decomposes into a
3-dim subspace spanned by $\{x_1,x_2,x_3\}$ and the remaining 2-dim
subspace which $\{x_4,x_5\}$ spans.\ $\vr\equiv (x_1,x_2,x_3)$ denotes
the coordinate in the 3-dim subspace.\
$\vrho_X$ denotes the coordinate in the 4-dimensional subspace
orthogonal to the $x_4$-direction and likewise $\vrho_Y$ the coordinate
in the 4-dim subspace orthogonal to the $x_5$-direction. In the last
line, we have chosen a general vector $\zeta\in (x_4,x_5)$ in the
45-plane and denoted the 4-dim subspace orthogonal to $\zeta$ by
the coordinates $\vrho_\zeta$.

Now we introduce $\zeta^1,\, \zeta^2 \in (x_4,x_5)$ as two
unit vectors in the 45-subspace in the 5-dim worldvolume, and  
define\ (with $a,..,e$ only spatial indices $\{1,2,3,4,5\}$)
\bea\label{def-tildeH}
\zeta^1,\, \zeta^2 \in (x_4,x_5) :\qquad
\tilde{\mathcal{H}}^{ab} \,=\, \frac 16 \epsilon^{abcde} H_{cde}\ ,\ && \
|\tilde{\mathcal{H}}|^2 = \tilde{\mathcal{H}}^{ab} \tilde{\mathcal{H}}^{cd}
\delta_{ac} \delta_{bd}\ ,\qquad \nn\\
|\del X|^2=\delta_{ab}\del_aX\del_bX\ ,\ && \
\partial_{[a} X \zeta^{1}_{b]} = \frac 12 \left( \partial_{a} X \zeta^{1}_{b}
\,-\, \partial_{b} X \zeta^{1}_{a}\right)\ .\quad
\eea
Since these are all spatial indices with a flat metric, we will not
distinguish up/down indices.

The presence of precisely two $\zeta$-vectors dovetails with including
only two of the five transverse scalars to the M5-brane. This
corresponds to including only two M2-branes ending on the M5-brane:
in an essential sense, this is incomplete from the point of view of
M5-M2 systems. However restricting to this, which reflects the decomposition
(\ref{notationrhozeta}), is adequate from the point of view of uplifts of
D3-brane configurations describing string webs. In other words, this
restriction implicitly encodes this subspace of $M\ra IIB$ dualities
obtained from compactifications of M-theory on $T^2\equiv (x_4,x_5)$.\
With this in mind, the energy functional for static configurations in
a single free M5-brane theory comprising the self-dual 3-form field
strength and two transverse scalars is 
\be\label{freelimEngy}
\mathcal{E} = \frac12 \left( |\partial X|^2 + |\partial Y|^2
+ \frac{1}2|\tilde{\mathcal{H}}|^2 \right) .\qquad
\ee
The sum here is over only spatial indices in ${\tilde H}$\,: this
incorporates the fact that $H$ is a self-dual 3-form field strength
with $H=\star_6 H$.
This can be abstracted from a superbrane formulation as discussed in
\cite{Townsend:1997wg}\ (see also \eg\
\cite{Bergshoeff:1998vx,Gauntlett:1997ss,Gauntlett:1999xz}). We will
find it adequate to focus on the minimal low energy point of view of
a single free M5-brane and its field content here, obtained by
retaining only the lowest mass dimension terms. The action for an
M5-brane with tension $T_6\sim {1\over l_{11}^6}$\ (with $l_{11}$
the 11-dim Planck length) \cite{Polchinski:1998rr} is written
in terms of the M5-brane coordinates $x^I$ in transverse space: these
are redefined to M5-brane worldvolume field theory scalars as
\be\label{M5-XI-fieldthLim}
{1\over l_{11}^6} \int d^6s\ (\del x^I)^2\ \equiv\ \int d^6s\ (\del X^I)^2\ ;
\qquad\quad X^I\equiv {1\over l_{11}^3}\,x^I\ ,
\ee
so the transverse scalar fields $X^I\equiv X, Y, \ldots$ are manifestly
dimension-2, and are the relevant observables in the field theory
regime.  The 2-form $B_{\mu\nu}$ has mass dimension-2 (so the Wilson
surface operator $\int B$ is dimensionless) and the scalars have mass
dimension-2, so all terms in ${\cal E}$ are dimension-6.

Upon Bogomolny completion of squares using the above $\zeta$-decomposition,
as in fact anticipated in \cite{Gauntlett:1999xz} (see
Appendix~\ref{AppA:identities} for some useful details), this gives\
\bea\label{completiongen}
&& {\cal E}\, =\, \left| \partial_{[a} X \zeta^{1}_{b]} \,+\, \partial_{[a} Y \zeta^{2}_{b]} \,-\, \frac 12 \delta_{ac} \delta_{bd} \tilde{\mathcal{H}}^{cd} \right|^2 \,+\, \frac 12 \left( \zeta^1_a \partial_a X + \zeta^2_a \partial_a Y \right)^2
\\ [1mm]
&& \qquad
+\, \partial_a X \tilde{\mathcal{H}}^{ab} \zeta^{1}_b \,+\, \partial_a Y \tilde{\mathcal{H}}^{ab} \zeta^{2}_b\ 
-\, \sum_{a \ne b} \zeta^{1}_a  \zeta^{2}_b \left( \partial_a X \partial_b Y - \partial_b X \partial_a Y\right) -\ \left( \zeta^{1} \cdot \zeta^{2} \right) \left( \partial X \cdot \partial Y \right)\quad \nn
\eea
Then the energy functional is minimized when the BPS bound equations 
\begin{align}\label{genBogo}
\partial_{[a} X \zeta^{1}_{b]} \,+\, \partial_{[a} Y \zeta^{2}_{b]} \,-\, \frac 12 \delta_{ac} \delta_{bd} \tilde{\mathcal{H}}^{cd} = 0\ , & \cr 
\zeta^1_a \partial_a X + \zeta^2_a \partial_a Y = 0 \ , &
\end{align}
are satisfied.\ We will study these equations and the resulting
configurations alongwith appropriate boundary conditions in what
follows.  The above essentially ignores the last two sets of terms in
(\ref{completiongen}): we will argue later in sec.~\ref{sec:tension},
sec.~\ref{sec:Bog1str-os} and sec.~\ref{tensionnon-ortho} that these
terms in fact give negligible contribution in the limit of approaching
the wall of marginal stability, and thereby lead to a self-consistent
energy minimization via the Bogomolny completion.

The 3-form field strength and its equation of motion 
can be recast giving
\be\label{EoMHmunu}
H_{abc} = \epsilon_{abcde} \tilde{\mathcal{H}}^{de}\quad\ra\quad
d\star H = 0\qquad\ra\qquad
\del_a \tilde{\mathcal{H}}^{ab} = 0 \ .
\ee
Thus
\be\label{EoMXY}
\sum_{\substack{a \\ a \ne b}} \partial_{a} ( \partial_{a} X \zeta^{1}_{b} - \partial_{b} X \zeta^{1}_{a})\,+\,  \sum_{\substack{a \\ a \ne b}} \partial_{a} ( \partial_{a} Y \zeta^{2}_{b} - \partial_{b} Y \zeta^{2}_{a})= 0 \,.
\ee
Imposing line source conditions along $\zeta^1, \zeta^2$ directions
leads to harmonicity in 4-dim subspaces orthogonal to the $\zeta$-vectors,
and thereby to harmonic field configurations transverse to string
soliton line sources stretched along $\zeta^1, \zeta^2$,
roughly ``superposing'' the two string solitons as we will describe.


It turns out that these are best described in component form, which we
will now proceed to do. In component form ($a=1,2,3$), we define
\be\label{HtildeH-comps}
E^a={\tilde {\cal H}}^{a4}\,,\qquad B^a={\tilde {\cal H}}^{a5}\,,\qquad
\Pi={\tilde {\cal H}}^{45}\,;\qquad
\vE=E^a{\hat x}^a\,,\quad \vB=B^a{\hat x}^a\,.
\ee
It is consistent to set\
$K^a =  \frac 12 \epsilon^{abc} \tilde{\mathcal{H}}^{bc} = 0$. In terms
of the 3-form $H$, these components are\ \
$B^a = \frac12 \epsilon^{abc} H_{bc4} , \ \
E^a = \frac12 \epsilon^{abc} H_{bc5} ,\ \ \Pi = H_{123} ,\ \
K^a = H_{a45}$.

Taking $\zeta^1,\ \zeta^2$ to lie in the $\{x_4, x_5\}$-plane (\ie\ 
no components in $\{1,2,3\}$), the BPS equations (\ref{genBogo}) in
component form become
\bea\label{BPScomp}
E_a = \del_a (X\zeta^1_4+Y\zeta^2_4)\,,\quad\
B_a = \del_a (X\zeta^1_5+Y\zeta^2_5)\,,\quad\ 
\Pi = \del_4(X\zeta^1_5+Y\zeta^2_5) - \del_5(X\zeta^1_4+Y\zeta^2_4)\,,&
\!\!\!\!
\nn\\ [2mm]
(\zeta^1_4\del_4+\zeta^1_5\del_5)X + (\zeta^1_4\del_4+\zeta^1_5\del_5)Y = 
\del_4(X\zeta^1_4+Y\zeta^2_4) + \del_5(X\zeta^1_5+Y\zeta^2_5) = 0\,.\qquad &
\eea
Defining appropriate $X', Y'$, leads to somewhat simpler expressions:
\bea\label{genBPS-X'Y'}
&& \qquad\qquad\qquad X' \equiv X\zeta^1_4+Y\zeta^2_4\ ,\qquad
Y' \equiv X\zeta^1_5+Y\zeta^2_5\ ,\nn\\ [1mm]
&& \vE=\nabla X'\ ,\qquad \vB=\nabla Y'\ ,\qquad \Pi=\del_4Y'-\del_5X'\ ,
\qquad \del_4X'+\del_5Y'=0\ .
\eea
The equations in this form were also obtained via supersymmetry in
\cite{Gauntlett:1999aw}. Our analysis of the resulting field configurations
is motivated by wall-crossing phenomena and string webs via D3-brane
prongs: as we will see, the map between $X', Y'$ and $X, Y$, involving
arbitrary vectors $\zeta^{1,2}$ in the $(x^4,x^5)$-plane  allows for
constructing string-string bound states for arbitrary charges.\
The Bogomolny completion (\ref{completiongen}) can be written as\ \
\bea\label{BogomolnyCompnnts}
&& {\cal E} = \frac12\big|\vE-\nabla X'\big|^2
+ \frac12\big|\vB-\nabla Y'\big|^2 + \vE\cdot\nabla X' + \vB\cdot\nabla Y'
\nn\\ [1mm]
&& \qquad\quad
+\ {1\over 2} (\zeta^1_a\del_aX + \zeta^2_a\del_aY)^2
+ \frac12\big|\Pi+\del_5X'-\del_4Y'\big|^2
+ \Pi (\del_4Y'-\del_5X')  \nn\\ [1mm]
&& \qquad\qquad\quad
-\, \sum_{a \ne b} \zeta^{1}_a  \zeta^{2}_b \left( \partial_a X \partial_b Y - \partial_b X \partial_a Y\right) - \left( \zeta^{1} \cdot \zeta^{2} \right) \left( \partial X \cdot \partial Y \right)
\eea
The simplest case with two orthogonal vectors\
$\zeta^1=(1,0),\ \zeta^2=(0,1)$,\ gives\ $X'=X,\ Y'=Y$.

Taking further derivatives of the last condition in (\ref{genBPS-X'Y'})
gives
\be
\del_4^2X'+\del_4\del_5Y'=0\ ,\qquad \del_4\del_5X'+\del_5^2Y'=0\ .
\ee
The equations of motion then give
\bea\label{delatildeHab}
&& \del_a{\tilde H}^{a4}=0:\qquad \nabla\cdot\vE - \del_5\Pi
= 0 = \nabla_i^2X'+\del_4^2X'+\del_5^2X'\ ,
\nn\\
&& \del_a{\tilde H}^{a5}=0:\qquad \nabla\cdot\vB + \del_4\Pi
= 0 = \nabla_i^2Y'+\del_4^2Y'+\del_5^2Y'\ .
\eea
In other words, $X', Y'$ are harmonic in the 5-dim spatial M5-worldvolume.
Since $X, Y$ are linear combinations of $X', Y'$, this implies that 
$X, Y$ are also similarly harmonic.


We have in mind two M2-branes ending on the M5-brane: each M2-brane
endline defines a self-dual string soliton in the M5-brane. So we have
two string solitons stretching along the $\zeta^1$ and $\zeta^2$
directions, which we restrict to lie in the $\{x_4,x_5\}$ plane but
take as general otherwise. This suggests that the two scalar field
configurations enjoy translation invariance along $\zeta^1$ and
$\zeta^2$ directions respectively.\ In the M5-brane effective field
theory this implies that the transverse scalars obey ``line source''
conditions. One might imagine imposing these line source conditions on
$X', Y'$, but for various technical reasons, it turns out more natural
to impose 
\be\label{lineSrcXY}
\zeta^1_a\del_a X = 0\ ,\qquad \zeta^2_a\del_a Y = 0\ ,
\ee
on $X, Y$. Thus
\be\label{XYharmonicity}
\nabla_i^2X+(\zeta^{1\,\perp}_m\del_m)^2X=0\ ,\qquad
\nabla_i^2Y+(\zeta^{2\,\perp}_m\del_m)^2Y=0\ ,
\ee
\ie\ $X, Y$ are harmonic in 4-dim subspaces transverse to $\zeta^1$
and $\zeta^2$ respectively. This follows from the 5-space harmonicity
of $X, Y$ upon imposing the line source conditions.\
So
\be\label{XYbasisConfig}
X={c_X\over |\vrho_{\zeta^1}-\vrho_{\zeta^1}^0|^2}-X_0 ,\qquad
Y={c_Y\over |\vrho_{\zeta^2}-\vrho_{\zeta^2}^0|^2}-Y_0 ,
\ee
using the notation (\ref{notationrhozeta}), with $c_X, c_Y$ being
arbitrary coefficients here. We then use (\ref{genBPS-X'Y'}) to write
the field configurations/states.  Then (\ref{genBPS-X'Y'}) shows that
the $X', Y'$ scalars being arbitrary linear combinations of the
``basis'' fields $X, Y$\ (\ref{XYbasisConfig}) give spike legs at
general angles.

We take the two strings to be located at $\vr=\vr_1$ and $\vr=\vr_2$,
\ie\ separated in the 3-space directions, and stretched along lines
in the $(x^4,x^5)$-plane in the $\zeta^1$ and $\zeta^2$ directions. By
using overall translation invariance along the $x^4,x^5$-directions, we
can always orient the strings to pass through the origin $(x^4,x^5)=(0,0)$.
Thus, using (\ref{notationrhozeta}) we can parametrize the coordinates
of the two strings as
\be\label{gen-stringCoords}
\vrho_1 = (\vr,x^4,x^5)_1 = (\vr_1,x_1\zeta_1^4,x_1\zeta_1^5) ,\qquad
\vrho_2 = (\vr,x^4,x^5)_2 = (\vr_2,x_2\zeta_2^4,x_2\zeta_2^5) ,\qquad\ \ 
\forall\ \ x_1, x_2 .
\ee
Thus the closest point of approach is at $(x^4,x^5)=(0,0)$, the
separation being $|\vr_1-\vr_2|$\ (see Figure~\ref{m5webprongs} (c)
for a specific case discussed later).
Now we impose boundary conditions mapping the worldvolume core
locations and moduli space: it might seem natural to impose moduli
boundary conditions all along the string spatial extents, \ie\ for
$\vrho_{\zeta^1}\ra\vrho_{\zeta^1}^0$ and $\vrho_{\zeta^2}\ra\vrho_{\zeta^2}^0$,
but this turns out to be overconstraining at generic points on the
M5-vacuum moduli space. Instead we impose only the weaker boundary
conditions below, at the closest approach between the two string
solitons:
\bea\label{bc-gen}
\vrho\ra (\vr_1,0,0)\,: && (X,Y)\ra (X_1,0)\quad \Rightarrow\qquad
(X', Y') \ra (X_1\zeta_4^1, X_1\zeta_5^1)\ ,\nn\\
\vrho\ra (\vr_2,0,0)\,: && (X,Y)\ra (0,Y_1)\quad\ \Rightarrow\qquad
(X', Y') \ra (Y_1\zeta_4^2, Y_1\zeta_5^2)\ .
\eea
In what follows, we will analyse the consequences of these boundary
conditions on the field configurations and the resulting M5-brane
bending geometry.

The boundary conditions suggest that $X', Y'$ lead to spike legs
at angles defined by $\zeta^1$ and $\zeta^2$ which being general imply
that general prong shapes are captured by the field
configurations (\ref{genBPS-X'Y'}) in terms of $X', Y'$. Thus we take
$X', Y'$ as the generic transverse scalars to the M5-brane. We will
now illustrate this with several examples.  The simplest case $(1,0),
(0,1)$ has $X'=X,\ Y'=Y$, which we will discuss in detail next.

\section{The simplest M5-M2 BPS states}\label{sec:m5m2}

A simple family of states is obtained when the two M2-branes end on
the M5-brane on orthogonal string solitons.  Then the $\zeta$-vectors
are orthogonal: without loss of generality we can take them to be
\be\label{zeta101zeta210}
\zeta^1 = {\hat x}^4 \equiv (1,0)\ ,\qquad
\zeta^2 = {\hat x}^5 \equiv (0,1)\ .
\ee
Then (\ref{genBPS-X'Y'}) and (\ref{genBogo}) give
\be\label{Bogo}
X'=X\ ,\quad Y'=Y\ ;\qquad
\vE = \nabla X\ ,\qquad \vB = \nabla Y\ ,\qquad 
\Pi = \del_4Y - \del_5X\ .
\ee
The line source conditions (\ref{lineSrcXY}) become
\be\label{del5X=0=del4Y}
\zeta^1\cdot\del X = \del_4X=0\ , \qquad \zeta^2\cdot\del Y = \del_5Y=0\ .
\ee
The 3-form field equations (\ref{delatildeHab}) alongwith
(\ref{del5X=0=del4Y}) now give
\be
\nabla_i^2X+\del_5^2X = 0\ ,\qquad \nabla_i^2Y+\del_4^2Y = 0\ ,
\ee
so the scalars $X, Y$ are harmonic in 4-subspaces orthogonal to $x_4$
and $x_5$ respectively.


\subsection{A single string soliton: the M5-brane spike}

\begin{figure}[h]
\bc\includegraphics[width=35pc]{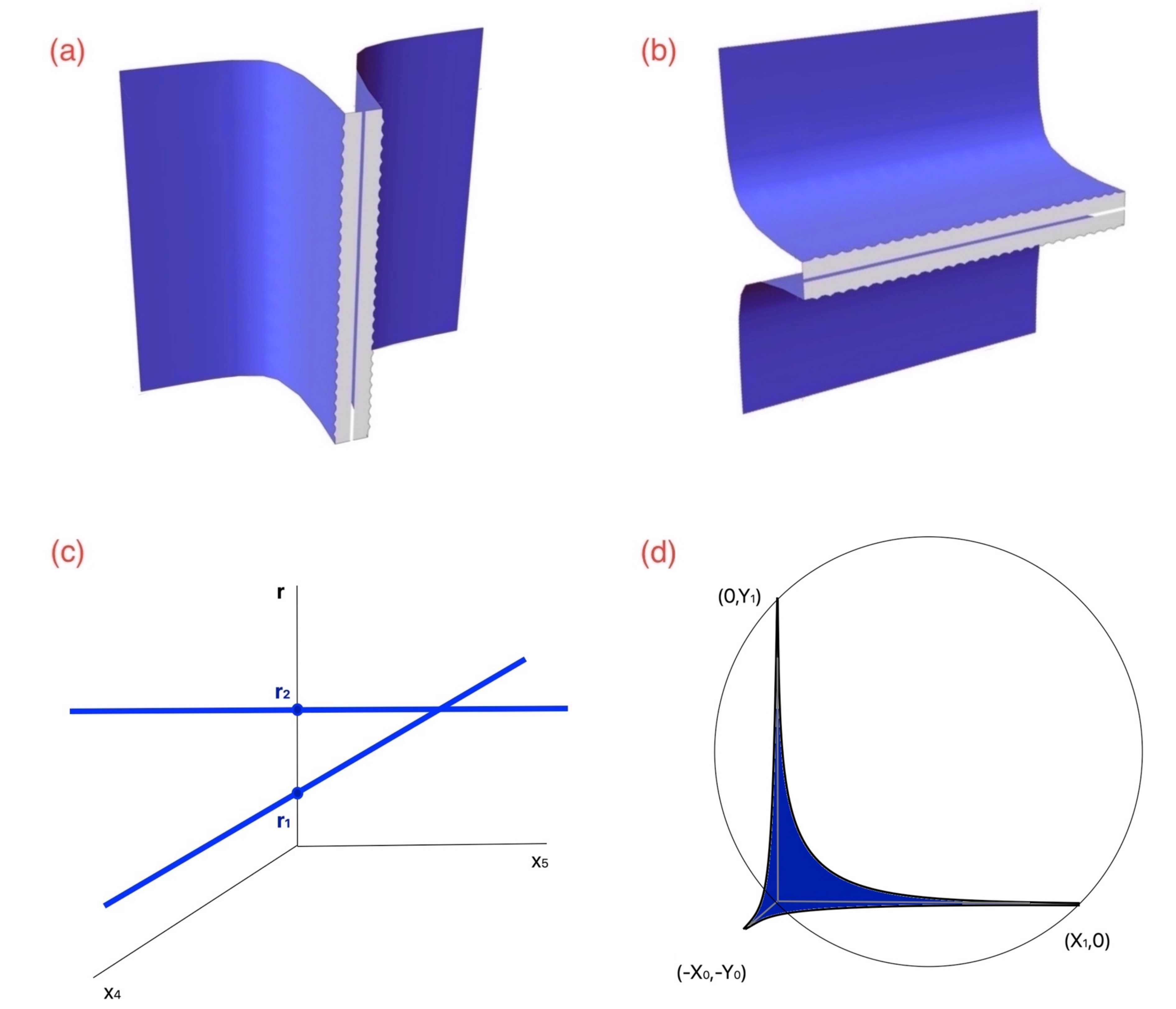}\ec
\caption{{ \label{m5webprongs}
    \footnotesize (a) and (b) depict the M5-brane ridge-shaped spike
    deformations $X\sim {1\over\vrho_X^2}$ and $Y\sim
    {1\over\vrho_Y^2}$ along the transverse $X$- and $Y$-directions
    respectively (the other worldvolume directions are
    suppressed). Fig.(c) shows the string soliton locations
    $\vrho_1=(\vr_1,x_1,0)$ and $\vrho_2=(\vr_2,0,x_2)$ in the
    M5-worldvolume (the strings are orthogonal, stretching along the
    $x^4$ and $x^5$ axes): the vertical axis represents the 3-dim
    subspace $\vr$. The M5-brane prong is roughly a superposition of
    spikes (a), (b).  Fig.(d) shows the shape in $(X,Y)$-transverse
    space of the M5 prong stretching out from the M5-brane at
    $(-X_0,-Y_0)$ (with individual spike-legs along $X$ and $Y$), in
    the limit $X_0, Y_0 \ll X_1, Y_1$, near the wall of marginal
    stability (circle).
}}
\end{figure}

Consider first a single M2-brane ending on the M5-brane giving a
self-dual string soliton stretched along say the $x^4$-direction so
$\zeta_X=(1,0)$.\ The field configuration and boundary condition are
\bea\label{M5spike}
&&  X = {q_e\over |\vrho-\vrho_{0}|^2} - X_0
= {q_e\over |\vrho_X-\vrho_{X,0}|^2} - X_0   
\equiv {q_e\over (\vr-\vr_0)^2+(x^5)^2} - X_0\,, \nn\\ [1mm]
&& {\tilde H}^{a4}{\hat x}^a = \vE = \nabla X\,,\qquad
       {\tilde H}^{45} = \Pi = -\del_5X\,;\qquad\quad 
       \zeta_X=(1,0):\quad \vrho_X=(\vr,x^5)\,.\qquad\ \
\eea
The first expression for $X$ is written in terms of the 5-dim coordinate
$\vrho$ while the second expression specifies the fact that $X$ is harmonic
in the 4-dim subspace transverse to the $x_4$-direction $\zeta_X$ that
the string soliton stretches along (the last expression makes this
explicit).
We see that there is a ${1\over \rho^2}$ divergence in $X(\vrho)$ at
the location of the string, parametrized as $\vrho=(\vr_0,x^4,0)$ using
(\ref{gen-stringCoords}):
the M2-brane ending on the M5-brane creates an infinitely long spike
on the M5-brane at the worldvolume string location where the
M2-brane intersects the M5-brane
\cite{Howe:1997ue,Gauntlett:1997ss}. The fact that the M2-brane
extends along the $x_4$ direction means that the ``spike'' is shaped
like a ``ridge'' with the ``flat ridge-top'' stretching along $x_4$ in
the M5-brane worldvolume, as reviewed in \cite{Townsend:1997wg}:
Figure~\ref{m5webprongs} (a) (or (b)) shows this ridge-shaped
M5-brane ``spike''.\ 
We can now impose a boundary condition on the scalar $X$,
\be\label{bc-M5spike}
\vrho\ra\vrho_{X,0}:\qquad X\ra X_1\ . 
\ee
This represents a long distance cutoff in the $X$-space transverse to
the M5-brane, which translates to a short distance cutoff around the
string soliton core: now the spike is of finite length. Later we will
embed these scalar configurations into the theory of multiple
M5-branes separated (``Higgsed'') and the cutoff will translate to the
distance between the M5-branes.

As $\vrho\ra\infty$, we see that $X\ra -X_0$\,: \ie\ far from the
string soliton core, the M5-brane deformation becomes vanishingly
small and the scalar $X$ relaxes to its modulus value $-X_0$
characterizing the M5-brane vacuum.
Now the boundary condition at $\vrho_X^0$ above gives
\be\label{al00-X}
q_e \al_{00} - X_0 \sim X_1\ \quad\Rightarrow\quad\ 
{1\over\al_{00}} \equiv \lim_{\vrho\ra\vrho_0^+} |\vrho-\vrho_0|^2 =
(\vr_c-\vr_0)^2+(x^5_c)^2\sim {q_e\over X_0+X_1}\ ,
\ee
where $\vrho_0^+$ means that $\vrho$ approaches $\vrho_0$ but is cut
off at $\vrho_c=(\vr_c,x^5_c)$\,. This leads to a finite ``core size'':
roughly speaking, we obtain a thickened (fuzzy) cylindrical tube. Thus
the cross-sectional size, or thickness, of the string core is given by
the inverse length of the M5-brane spike in transverse (moduli) space.

For what follows we will find it useful to write the configuration
(\ref{M5spike}) more generally as
\be\label{M5spike-zeta}
X = {q_e\over |\vrho_\zeta-\vrho_{\zeta,0}|^2} - X_0
= {q_e\over (\vr-\vr_0)^2 + (\zeta^\perp\cdot(\vx-\vx_0))^2} - X_0\ ,
\qquad \vx\cdot\zeta=0\ ,
\ee
where $\zeta^\perp\in (x_4,x_5)$ is a unit vector orthogonal to $\zeta$.
For instance $\zeta_X=(1,0)$ has $\zeta_X^\perp=(0,1)$ recovering the
field configuration in (\ref{M5spike}).\
It is also instructive to note that we can likewise construct similar
M5-brane spike configurations carrying generic dyonic charge $(q_e,q_m)$:
consider
\bea\label{M5spikeDyon}
\zeta^1 = {1\over\sqrt{q_e^2+q_m^2}} (q_e,q_m)\equiv \gamma (q_e,q_m)\ , 
\qquad \zeta^2=0\ ,\qquad\qquad\qquad & \nn\\ [1mm]
X' = {q_e\over (\vr-\vr_0)^2 + (\zeta^\perp\cdot(\vx-\vx_0))^2} - X_0'\,,
\ \quad\
Y' = {q_m\over (\vr-\vr_0)^2 + (\zeta^\perp\cdot(\vx-\vx_0))^2} - Y_0'\,, &
\nn\\ [2mm]
{\tilde H}^{a4}{\hat x}^a = \vE = \nabla X'\,,\quad\ \
{\tilde H}^{a5}{\hat x}^a = \vB = \nabla Y'\,,\quad\ \
{\tilde H}^{45} = \Pi = \del_4Y'-\del_5X'\,.\quad\ &
\eea
We see that $(\vE,\vB) \propto (q_e,q_m)$. We have constructed
$X', Y'$ using a single ``basis'' field $X$ with $Y=0$: the coefficients
in (\ref{XYbasisConfig}) are $c_X=1/\gamma,\ c_Y=0$\,. This can be
thought of as an electric string ``rotated'' through a single 
nontrivial $\zeta^1$-vector to give a dyonic string soliton.

\medskip

\bc
\ec

\subsection{String soliton bound states: M5-brane prongs}

Now we describe field configurations given by two scalars on a single
M5-brane: the M5-brane worldvolume deformation, roughly speaking, is
now two-pronged.  In an essential sense, this arises by trying to
superpose two string solitons, each being the endline of an M2-brane
ending on the M5-brane. These are stretched along two distinct
non-parallel M5-worldvolume spatial directions, creating a
superposition of two M5-brane spikes, one along the $X'$-direction and
the other along the $Y'$-direction. Imposing boundary conditions on
the scalar moduli in the vicinity of the string ``cores'' ensures that
the resulting configuration is a molecule-like bound state of the two
non-parallel string solitons with fixed transverse separation along
the $\vr$-direction, as we will see. Intuition for these boundary
conditions arises from string web states in \Nf\ SYM theories arising
on D3-branes (a brief recap appears in Appendix~\ref{sec:D3prongs}).
We will call the resulting M5-brane deformations as M5-brane
``prongs''.



\noindent \underline{\emph{$(1,1)-(1,0)-(0,1)$ prong}}:\ \
The simplest such configuration is described by the $(1,1)-(1,0)-(0,1)$
state: in the D3-brane limit, this is described by the bound state
of a fundamental $F1$- and $D1$-string ending on a D3-brane. This can
be viewed as M-theory with the 45-directions compactified on a torus
$T^2_{45}$:\ in the uncompactified M-theory description we have an
M5-brane with two M2-branes stretched along say the $x_6$- and
$x_7$-directions, and ending on lines stretching along the $x_4$- and
$x_5$-directions in the M5-brane worldvolume: upon compactification
the M5-brane becomes the D3-brane while the M2-branes stretched along
$x_4$ and $x_5$ become respectively the F1- and D1-string.


The M5-M2 brane orientations are (suppressing directions that do not
enter)
\be\label{M5M2geom}
\bA{ccccccccc} & 0 & 1 & 2 & 3 & 4 & 5 & 6 & 7 \\
D3\ \  & \bullet & \bullet & \bullet & \bullet & & & & \\
F1\ \  & \bullet & & & &  & & \bullet & \\
D1\ \  & \bullet & & & & & & & \bullet \\
\eA\ \ldots
\quad\ \ \longleftrightarrow\quad\ \
\bA{ccccccccc} & 0 & 1 & 2 & 3 & 4 & 5 & 6 & 7 \\
M5\ \  & \bullet & \bullet & \bullet & \bullet & \bullet & \bullet & & \\
M2\ \  & \bullet & & & & \bullet & & \bullet & \\
M2\ \  & \bullet & & & & & \bullet & & \bullet \\
\eA\ \ldots\
\ee
In the D3-brane effective field theory, this is described as a bound
state representing a $(1,1)$-charge molecule, with constituent charges
$(1,0)$ and $(0,1)$: geometrically this can be described by a ``prong
web'' obtained by deformations of the D3-brane worldvolume with two
scalars.  We expect and are looking for a similar prong-web
description in the uncompactified M-theory, with two scalars
characterizing nontrivial M5-brane worldvolume deformations
representing a nontrivial bound state of the two string solitons
resulting from the end-lines of the two M2-branes ending on the
M5-brane.

We will take the strings to be stretched along the $x^4$ and $x^5$
directions but separated in the $\vr$-directions\
(Figure~\ref{m5webprongs} (c)): the $\zeta$-vectors and string
coordinate parametrizations (\ref{gen-stringCoords}) are
\bea\label{1001-stringCoords}
\zeta_1=(1,0) :\qquad \vrho_1 = (\vr,x^4,x^5)_1 = (\vr_1,x_1,0)\quad
\forall\ x_1\,;\nn\\
\zeta_2=(0,1) :\qquad \vrho_2 = (\vr,x^4,x^5)_2 = (\vr_2,0,x_2)\quad
\forall\ x_2\,.
\eea
We have used the translation symmetry along the $x^5$ direction to
set the locations $x^5_1$ and $x^4_2$ of the string solitons to be
$x^5_1=0$ and $x^4_2=0$\,: in other words, we have taken the strings
to lie along the $x^4$ and $x^5$ axes, without loss of generality.
We expect this system can be described by the scalar field
configurations (using the notation (\ref{notationrhozeta}))
\bea\label{m5m2-111001}
X = {q_e\over |\vrho_X-\vrho_X^1|^2} - X_0
= {q_e\over (\vr-\vr_1)^2 + (x^5)^2} - X_0\ , &&\ \ 
\qquad\  \vrho_X\cdot\zeta_1=0\ ,    
\nn\\ [1mm]
Y = {q_m\over |\vrho_Y-\vrho_Y^2|^2} - X_0
= {q_m\over (\vr-\vr_2)^2 + (x^4)^2} - Y_0\ , &&\ \
\qquad\ \vrho_Y\cdot\zeta_2=0\ ,
\nn\\ [2mm]
{\tilde H}^{a4}{\hat x}^a = \vE = \nabla X\,,\qquad
{\tilde H}^{a5}{\hat x}^a = \vB = \nabla Y\,,\quad &&
{\tilde H}^{45} = \Pi = \del_4Y-\del_5X\,.\
\eea
These are roughly the superposition of an electric spike
(\ref{M5spike}) along $X$ and a magnetic one along $Y$, as shown 
in Figure~\ref{m5webprongs} (a) and (b).\ (Note that electric/magnetic
here are simply labels for $4/5$-directions of the strings which
are self-dual solitons in 6-dimensions; upon 45-compactification, these
will map to electric/magnetic point charges in 4-dimensions.)\
They can be seen to be a valid solution to the BPS bound equations
(\ref{genBPS-X'Y'}) alongwith the line source conditions
(\ref{lineSrcXY}) with the $\zeta$-vectors above (and thereby also
satisfy harmonicity (\ref{XYharmonicity}) in the appropriate transverse
4-subspaces).\
We can rewrite the field configurations (\ref{m5m2-111001}) as
\be\label{m5m2-111001'}
X = {q_e\over (\vr-\vr_1)^2 + (\zeta_1^\perp\cdot\vx)^2} - X_0\ ,
\qquad
Y = {q_m\over (\vr-\vr_2)^2 + (\zeta_2^\perp\cdot\vx)^2} - Y_0\ .  
\ee
The distance elements here correspond to the distance in the
subspace transverse to the string soliton, which thus is conveniently
expressed using the unit normals $\zeta_1^\perp,\ \zeta_2^\perp$ to the
string orientations $\zeta_1, \zeta_2$\ (similar to (\ref{M5spike-zeta})).

We will now examine the M5-brane deformation geometry described by
these field configurations. Firstly, as we go out to infinity on the
M5-brane, we expect the deformations (\ref{m5m2-111001}) become
vanishingly small, \ie\
\be
\vrho\ra\infty:\qquad (X,Y) \ra (-X_0, -Y_0)\ ,
\ee
which is the location of the M5-brane. The strings extend in one of the
$x^4,x^5$-directions: if we go out to infinity in the 3-space directions,
the deformations (\ref{m5m2-111001}) scale as
\be\label{111001-rInfty}
\vr\ra\infty,\ \ x^4, x^5\ {\rm finite}\,:\qquad
X\sim {q_e\over |\vr|^2}-X_0\ ,\quad Y\sim {q_m\over |\vr|^2}-Y_0\ .
\ee
Thus when
\be\label{moduliConstr-111001}
{1\over q_e} X_0 = {1\over q_m} Y_0\ ,
\ee
the M5-brane deformation in this asymptotic regime traces a line
along $(1,1)$ in the $X,Y$-space transverse to the
M5-brane in $q_e,q_m$-units. Likewise the electric and magnetic field
components in (\ref{m5m2-111001}) scale as $(\vE,\vB)\propto (1,1)$,
thereby encoding charge $(1,1)$. This relation
(\ref{moduliConstr-111001}) on the asymptotic moduli also arises from
an understanding of the effective tension of these configurations, as
we will see in Sec.~\ref{sec:tension}.

Let us now study the scalar field configurations as we approach 
string$_1$ stretched along $\vrho_1=(\vr_1,x_1,0)$ in
\ref{1001-stringCoords}). We see that the $X$
scalar displays a spike as expected, similar to (\ref{M5spike}),
(\ref{al00-X}). Looking at both scalars, we find in this limit, 
\be\label{111001-Y:x4}
X \sim \lim_{\vrho\ra\vrho_1^+} {q_e\over |\vrho-\vrho_1|^2} - X_0\ \ra\ X_1\ ;
\qquad
Y \sim {q_m\over (\vr_1-\vr_2)^2 + (x_1)^2} - Y_0\ .
\ee
$X_1$ is the cutoff value of the divergent $X$ spike, similar to
(\ref{bc-M5spike}). The $Y$ scalar on the
other hand has no singularity anywhere. Since $x_1$ here labels the
coordinate along string$_1$, if we go out to large $x_1$ along the
string, the first term dies and we obtain $Y\ra -Y_0$. However at
$x_1=0$ we obtain
\be\label{111001-Y:x4=0}
x_1=0:\qquad Y\sim {q_m\over (\vr_1-\vr_2)^2} - Y_0\ .
\ee
Likewise as we approach string$_2$ stretched along $\vrho=(\vr_2,0,x_2)$
in \ref{1001-stringCoords}), we see that the $Y$ scalar displays a
spike as expected, similar to (\ref{M5spike}), (\ref{al00-X}),
but the $X$ scalar appears finite.
The scalars approach
\be\label{111001-X:x5}
X \sim {q_e\over (\vr_1-\vr_2)^2 + (x_2)^2} - X_0\ ;\qquad
Y \sim \lim_{\vrho\ra\vrho_2^+} {q_m\over |\vrho-\vrho_2|^2} - Y_0\ \ra\ Y_1\ .
\ee
$Y_1$ is the cutoff value of the $Y$ spike similar to
(\ref{bc-M5spike}), and $X$ exhibits
no singularity anywhere. $x_2$ here labels the coordinate along
string$_2$, so for large $x_2$ along the
string, the first term dies and we obtain $X\ra -X_0$. However at
$x_2=0$ we obtain
\be\label{111001-X:x5=0}
x_2=0:\qquad X\sim {q_e\over (\vr_1-\vr_2)^2} - X_0\ .
\ee
These points (\ref{111001-X:x5=0}), (\ref{111001-Y:x4=0}), are
special: they lie on the closest point of approach between the two
strings. The distance between a point on string$_1$ and any point on
string$_2$ is given by\ $(\vr_1-\vr_2)^2+(x_1)^2+(x_2)^2$ which is
minimum when $x_1=x_2=0$\,: projecting onto the $(x^4,x^5)$-plane,
this is simply the intersection of the lines along $(x_1,0)$ and $(0,x_2)$.

From similar bound state configurations arising from D3-brane prongs
(Appendix~\ref{sec:D3prongs}), we know that the separation
between the constituent charge cores is fixed and scales as the
inverse distance from the wall of marginal stability.
Using this intuition, let us imagine that the transverse
distance $|\vr_1-\vr_2|$ between the strings is fixed. It is then
consistent to impose from (\ref{111001-Y:x4=0}), (\ref{111001-X:x5=0}), 
that
\be\label{111001-M5wms}
{1\over (\vr_1-\vr_2)^2} = {1\over q_e} X_0 = {1\over q_m} Y_0\ .
\ee
Overall, this suggests the boundary conditions (\ref{bc-gen}),
\be\label{bc-111001}
\vrho\ra (\vr_1,0,0):\quad (X,Y)\ra (X_1,0)\ ;\qquad\quad
\vrho\ra (\vr_2,0,0):\quad (X,Y)\ra (0,Y_1)\ ,
\ee
imposed at the closest point of approach of the two orthogonal strings.
These boundary conditions are equivalent to fixing the transverse
separations between the strings as (\ref{111001-M5wms}).
As we have seen, the scalar moduli values vary in the various
asymptotic M5-brane worldvolume limits (\ref{111001-rInfty}),
(\ref{111001-Y:x4}), (\ref{111001-Y:x4=0}), (\ref{111001-X:x5}),
(\ref{111001-X:x5=0}): these variations, summarized in
Table~\ref{tableXYbndy21}, are consistent with (\ref{bc-111001}) and
follow from the field configurations (\ref{m5m2-111001}).
\begin{table}[H]
	\begin{center}
		\begin{tabular}{|c|c|c|}
			\hline
			$\quad(X, Y) \rightarrow (X_1, 0)\quad$ &  $\quad\text{}$ $\vr \rightarrow \vr_1 \quad$ & $\!\!\!\! x_1  = 0$ \cr
			\hline
			$(X, Y) \rightarrow (X_1, -Y_0)$ &  $\vr \rightarrow \vr_1$ & $ |x_1| \rightarrow \infty$  \cr
			\hline
			$(X, Y) \rightarrow (0, Y_1)$ &  $\vr \rightarrow \vr_2$ & $ x_2 = 0$ \cr
			\hline
			$(X, Y) \rightarrow (- X_0, Y_1)$ &  $\vr \rightarrow \vr_2$ & $|x_2| \rightarrow \infty$ \cr
			\hline
			$(X,Y) \rightarrow (-X_0, - Y_0)$ & $\vr$ $\rightarrow \infty$ &$|\vx|$ $\rightarrow \infty$  \cr
			\hline
		\end{tabular}
	\end{center}
	\caption{The $(X,Y)$ scalar moduli values
          at various regimes along the two strings.} 
	\label{tableXYbndy21}
\end{table}

Now, along with the transverse separations (\ref{111001-M5wms}), we
also obtain the inverse cross-sectional core sizes of the
two string solitons from (\ref{111001-Y:x4}), (\ref{111001-X:x5}),
\be\label{111001-coreSize}
\al_{ii} \equiv \lim_{\vrho\ra\vrho_i^{+}} {1\over |\vrho-\vrho_i|^2}\ ;
\qquad\quad
\al_{11} = {X_0+X_1\over q_e}\ ,\qquad \al_{22} = {Y_0+Y_1\over q_m}\ ,
\ee
similar to the cutoff core size in (\ref{al00-X}) for a single spike.

From (\ref{111001-M5wms}) and (\ref{111001-coreSize}), we see that
\be\label{111001-WMSlimit}
X_0, Y_0\ra 0:\qquad |\vr_1-\vr_2|\ra\infty\,,\qquad
\al_{11}^{-1}\sim {q_e\over X_1}\,,\quad \al_{22}^{-1} \sim {q_m\over Y_1}\ ,
\ee
\ie\ the constituent string solitons are infinitely separated in the 
3-space directions transverse to their extended directions, while
their core sizes $\al_{ii}^{-1}$ are finite (and small if $X_1, Y_1$
are large).
Thus the bound state becomes arbitrarily loosely bound, \ie\ it decays
and disappears from the spectrum.\ There is no physical bound state
solution with $(\vr_1-\vr_2)^2>0$ to (\ref{111001-M5wms}) for $X_0<0$\
(the other side of the wall).

From Table~\ref{tableXYbndy21} summarizing the moduli variations, we
note from (\ref{111001-X:x5}) that the M5-brane $X$-deformation along
string$_2$ varies from $X=0$ at $x_2=0$ to $X=-X_0$ at large $x_2$,
and likewise the M5-brane $Y$-deformation along string$_1$
interpolates from $Y=0$ at $x_1=0$ to $Y=-Y_0$ at large $x_1$. Thus
the M5-brane bending is nontrivial as we go out to infinity along the
strings. Note that this does not happen for a single string
soliton obtained from a single M2-brane ending on the M5-brane (there
are some qualitative parallels with the M5-brane descriptions of
\Nt\ theories \cite{Witten:1997sc}, although the brane bending is
less severe in the present discussions).
Overall however, in the limit of the M5-brane approaching the wall
of marginal stability
\be
X_0, Y_0\ll X_1,\ Y_1\ ,
\ee
this variation of $X, Y$ is negligible compared with the individual
spike legs which are much larger. Thus in this limit $X_0,
Y_0\ll X_1, Y_1$, the shape of the M5-brane in the transverse
$X,Y$-space resembles Figure~\ref{m5webprongs} (d): this refinement
of the scalar boundary conditions and their asymptotics is similar
to that in \cite{Argyres:2001pv} for D3-brane SYM theories (but with
additional intricacies due to the string-nature of the constituents).
There are also parallels in the qualitative picture of the decay
across the wall of marginal stability: wall-crossing amounts to the
state becoming arbitrarily loosely bound, and is qualitatively similar
to wall-crossing for black holes \cite{Denef:2000nb}. The limit
$X_0, Y_0\ra 0$ of approaching the wall of marginal stability (circle
in Fig.~\ref{m5webprongs} (d)) means the shortest leg approaches zero
size, \ie\ the M5-brane at $(-X_0,-Y_0)$ approaches $(0,0)$.

It is also instructive to note that far from both of the string
solitons, we obtain
\be
X \sim {q_e\over (\vr)^2 + (x_2)^2} - X_0\ ,\qquad
Y \sim {q_m\over (\vr)^2 + (x_1)^2} - Y_0\ ,
\ee
and the field configurations appear to describe two string solitons
intersecting at $(x_1,x_2)=(0,0)$. In other words the transverse
$\vr$-separation is
not discernable and the two M2-branes now appear to approximately
intersect each other on the M5-brane when viewed from afar.
Note that this appears slightly
different from the D3-brane limit where the two pointlike cores
appear to effectively coalesce: the string solitons being extended
objects along orthogonal directions remain orthogonal and extended.

Another interesting point, using the relation (\ref{M5-XI-fieldthLim})
between the worldvolume scalars in the field theory regime and the
M5-brane coordinates, is 
\be\label{111001-M5wms-2}
(\vr_1-\vr_2)^2 \sim {1\over X_0} \sim {l_{11}^3\over x_0}\ .
\ee
If we consider holding the M5-brane coordinate distance $x_0$ fixed,
taking $l_{11}\ra 0$ implies that $|\vr_1-\vr_2|\ra 0$ generically,
which suggests a junction-like configuration: the exception occurs
when $x_0=0$ beyond which point the bound state ceases to exist.
This appears to pertain to comments on M2-brane junctions in \eg\
\cite{Lee:2006gqa} and \cite{Gaiotto:2009hg}, as well as
\cite{Schwarz:1996bh}.
On the other hand, we have been describing the field theory regime,
which arises by taking $l_{11}\ra 0$ holding the ``field theory
distance'' $X_0$ fixed (essentially, as in \cite{Maldacena:1997re}).
In this limit we obtain a smooth description of the string soliton
molecule decaying at the wall of marginal stability as $X_0\ra 0$\
(with no ``junction'' per se). Something similar occurs for D3-branes
and string webs in the field theory regime, as discussed in
\cite{Argyres:2001pv,Argyres:2000xs}.

\subsubsection{M5-brane prong for $(n,m)-(n,0)-(0,m)$}

Along very similar lines, we can describe the $(n,m)-(n,0)-(0,m)$
M5-brane prong web
representing a charge-$(n,m)$ string soliton molecule, with constituent
string solitons of electric and magnetic charges $(n,0)$ and $(0,m)$.\ 
The $\zeta$-vectors are again orthogonal, as in (\ref{1001-stringCoords})
with $\zeta_1=(1,0),\ \zeta_2=(0,1)$, so the scalar field configurations
again satisfy $X'=X,\ Y'=Y$. This is a 2-parameter family of string
soliton bound states but the description is very similar overall to
the case we just described, with $n=1,\ m=1$, so we will be brief here.
The scalar field configurations are
\be\label{m5m2-nmn00m}
X = {n\,q_e\over (\vr-\vr_1)^2 + (x^5)^2} - X_0\ ,\qquad
Y = {m\,q_m\over (\vr-\vr_2)^2 + (x^4)^2} - Y_0\ ,
\ee
very similar to (\ref{m5m2-111001}).
The $(X,Y)$-moduli asymptotically, as $\vrho\ra\infty$, trace out a
line along $(n,m)$ in the moduli space when
\be\label{moduliConstr-nmn00m}
{m\over q_e} X_0 = {n\over q_m} Y_0\ .
\ee
Imposing the boundary conditions (\ref{bc-111001}) on the scalars
leads to 
\be\label{nmn00m-M5wms}
{1\over (\vr_1-\vr_2)^2} = {1\over n q_e} X_0 = {1\over m q_m} Y_0\ ,
\ee
as the transverse separation between the two orthogonal string
solitons in the M5-brane worldvolume.\ 
We then also obtain the inverse core sizes
\be
\al_{11} = {X_0+X_1\over n q_e}\ ,\qquad \al_{22} = {Y_0+Y_1\over m q_m}\ .
\ee
The asymptotic behaviour of the field configurations, as well as the
picture of the decay as the M5-brane approaches the wall of marginal
stability, is very similar to the $(1,1)-(1,0)-(0,1)$ case discussed
earlier.

\subsection{1-string tension, marginal stability and
  wall-crossing}\label{sec:tension}

We will now calculate the tension of the string soliton bound states
we have been describing. It is useful to note that the 
$\zeta$-vectors we have been using admit a rotation
by an overall angle $\varphi$ (which is just an overall rotation of
the $x^4$-$x^5$ axes): this gives a slightly more general representation
of the orthogonal $\zeta$-vectors in (\ref{1001-stringCoords}) as
\be\label{zetaOrtho-varphi}
\zeta^1 = \cos \varphi \, \hat{x}^4 - \sin \varphi \, \hat{x}^5\ ,\qquad
\zeta^2 =  \sin \varphi \, \hat{x}^4 + \cos \varphi \, \hat{x}^5\ .
\ee
while retaining the geometry and physics of the field configurations.
Then using (\ref{genBPS-X'Y'}), the Bogomolny completion
(\ref{BogomolnyCompnnts}) becomes\ (with $\zeta^1\cdot\zeta^2=0$)
\bea\label{BogomolnyCompnnts-2}
& {\cal E} = \frac12\big|\vE-\cos\varphi\nabla X-\sin\varphi\nabla Y\big|^2
+ \frac12\big|\vB+\sin\varphi\nabla X-\cos\varphi\nabla Y\big|^2 \nn\\ [1mm]
& +\ \frac12
\big|\Pi+\cos\varphi(\del_5X-\del_4Y)+\sin\varphi(\del_4X+\del_5Y)\big|^2
\nn\\
& +\ \frac12
\big( \sin\varphi(\del_5X-\del_4Y) - \cos\varphi (\del_4X+\del_5Y) \big)^2
\nn\\ 
& \qquad  +\ \vE\cdot(\cos\varphi\nabla X+\sin\varphi\nabla Y)
+ \vB\cdot(-\sin\varphi\nabla X+\cos\varphi\nabla Y) \nn\\ [1mm]
& \qquad -\ \Pi\big(\del_5(\cos\varphi X+\sin\varphi Y)
-\del_4 (-\sin\varphi X+\cos\varphi Y)\big)
\nn\\ [1mm]
& \qquad\ 
-\, (\del_4X\del_5Y-\del_5X\del_4Y)
\eea

We will now show that the cross-terms $\del_4X\del_5Y-\del_5X\del_4Y$
in (\ref{BogomolnyCompnnts-2}) upon integrating over 5-space have
negligible contribution on-shell. With nonzero $\varphi$, the
expressions for the $X$ and $Y$ fields using (\ref{m5m2-111001'}) with
(\ref{zetaOrtho-varphi}) become
\be\label{XandYvarphi}
X = \frac{q_e}{\left( \vr - \vr_1 \right)^2 + \left( \sin \varphi \, x^4 + \cos \varphi \, x^5 \right)^2} - X_0\,,  \quad\ \
Y = \frac{q_m}{\left( \vr - \vr_2 \right)^2 + \left( \cos \varphi \, x^4 - \sin \varphi \, x^5 \right)^2} - Y_0\,.
\ee
The cross term under the 5-space integral is
\be\label{crossintegral}
\int d^3 r dx^4 dx^5 \left( \del_4X\del_5Y-\del_5X\del_4Y \right)
\ee
It is convenient to consider an overall rotation by $\varphi$ in the
$x^4-x^5$ plane by defining
\be
x'^4 = \cos \varphi \, x^4  -  \sin \varphi \, x^5 \,,  \qquad
x'^5 = \sin \varphi \, x^4  +  \cos \varphi \, x^5 \,.
\ee
Then (\ref{XandYvarphi}) takes a simpler form
\be\label{XandYvarphi2}
X = \frac{q_e}{\left( \vr - \vr_1 \right)^2 + \left( \, x'^5\right)^2} - X_0\,,
\qquad
Y = \frac{q_m}{\left( \vr - \vr_2 \right)^2 + \left(  x'^4 \right)^2} - Y_0\,,
\ee
and the form of the cross-terms in (\ref{crossintegral}) remains the
same\footnote{Essentially these terms have the structure of the
  corresponding terms in the wedge-product $\int dX\wedge dY$ so they
  are manifestly unchanged under the coordinate transformation. This can
  also be checked using\ 
  $\partial_4 =\cos \varphi \frac{\partial}{\partial x'^4} + \sin \varphi \frac{\partial}{\partial x'^5}$ and $\partial_5 = - \sin \varphi \frac{\partial}{\partial x'^4} + \cos \varphi \frac{\partial}{\partial x'^5}$\ and evaluating
  explicitly.}
\be
\int d^3 r dx'^4 dx'^5 \left( \del_{4'}X\del_{5'}Y-\del_{5'}X\del_{4'}Y \right)
\ee
The first term $\del_{4'}X\del_{5'}Y$ vanishes and the second term
upon integration by parts gives
\begin{align}
-  \int d^3r dx'^4 \left[ X \partial_{4'} Y \Big|_{x'^5 = 0^+}^{\infty} \,+\, X \partial_{4'} Y \Big|^{x'^5 = 0^-}_{-\infty} \right]
\end{align}
In the above, we have introduced boundary values near $x^5 = 0$ value due
to string$_1$ when $\vec{r} = \vec{r}_1$. When $\vec{r} \ne \vec{r}_1$
the effect of doing this in the integral expression is negligible.

$\partial_{4'} Y$ can be factored out here since it is independent of
$x'^{5}$ and another integration by parts can be done here $w.r.t.$ $dx'^4$
giving
\begin{align}
-  \int d^3r  \left[ X  \Big|_{x'^5 = 0^+}^{\infty} \,+\, X  \Big|^{x'^5 = 0^-}_{-\infty} \right]  \left[ Y  \Big|_{x'^4 = 0^+}^{\infty} \,+\, Y  \Big|^{x'^4 = 0^-}_{-\infty} \right]
\end{align}
Since the $X$ and $Y$ fields do not depend on the signs of $x'^5$ and
$x'^4$ coordinates, we conclude that the square bracket terms vanish.
Note that this vanishing of these cross-terms (last line in
(\ref{BogomolnyCompnnts-2})) is on-shell, using (\ref{XandYvarphi2}).

Now, the terms in lines 4 and 5 in (\ref{BogomolnyCompnnts-2}) upon
integrating over 5-space can be recast as
\begin{align}
\int dx^4 \int d^3r dx^5  \, \tilde{\mathcal{H}}^{a4} \left( \cos\varphi \,
\partial_a X  + \sin \varphi \, \partial_a Y \right)
+ \int dx^5 \int d^3r dx^4 \, \tilde{\mathcal{H}}^{a5} \left( -\sin\varphi \,
\partial_a X  + \cos \varphi \, \partial_a Y \right) 
\end{align}
which upon doing the integration by parts and subsequently using the
equation of motion $\partial_a \tilde{\mathcal{H}}^{ab} = 0$ become 
\begin{align}
\int dx^4 \oint_{S^3} d\hat{s}^a \,\, \tilde{\mathcal{H}}^{a4} \left( \cos \varphi \,  X  \,+\, \sin \varphi \,  Y \right) \, + \,\int dx^5	\oint_{S^3} d\hat{s}^a \,\, \tilde{\mathcal{H}}^{a5} \left(  - \sin \varphi \, X  \,+\, \cos \varphi \,  Y \right) .
\end{align}
We are considering two strings with charges
\be
(Q_E^i,Q_B^i),\ i=1,2 :\qquad
Q_E^i=\oint_{S^3_{X,i}} {\tilde H}^{a4} ds^a\ ,\qquad
Q_B^i=\oint_{S^3_{Y,i}} {\tilde H}^{a5} ds^a\ ,
\ee
with scalar moduli values at their cores $(X'_i,Y'_i)$. At spatial
infinity far from both strings, charge conservation gives
$(Q_E^0,Q_B^0)=-\sum_i (Q_E^i,Q_B^i)$ and the moduli values are
$(-X_0,-Y_0)$.\ For a single string, the asymptotic boundary to
the 4-dim transverse space is $S^3$. For the combined configuration of
two strings we are discussing, integrating over 5-space requires
incorporating the appropriate $S^3_{X,i}$ and $S^3_{Y,i}$, giving
\be\label{extr-varphi}
L \sum_i\ \cos\varphi \big( (X_i+X_0)Q_E^i + (Y_i+Y_0)Q_B^i \big)\,
+\, \sin\varphi \big( (Y_i+Y_0)Q_E^i - (X_i+X_0)Q_B^i \big)\ ,
\ee
with $L\equiv \int dx^4 \sim \int dx^5$ the regulated length of both
strings. Extremizing w.r.t. $\varphi$ gives 
\be\label{T-bound}
T\equiv {M\over L} = \sqrt{ \big( (X_i+X_0)Q_E^i + (Y_i+Y_0)Q_B^i \big)^2
  + \big( (Y_i+Y_0)Q_E^i - (X_i+X_0)Q_B^i \big)^2 }
\ee
as the string tension obtained with the tightest bound, with
\be\label{varphiXY}
\tan\varphi = {(Y_i+Y_0)Q_E^i - (X_i+X_0)Q_B^i \over
  (X_i+X_0)Q_E^i + (Y_i+Y_0)Q_B^i}
\ee
This extremization analysis for these boundary terms cares only about
the transverse $(X,Y)$-space and the charges and is very similar to
that in the D3-brane case reviewed briefly in App.~\ref{sec:D3prongs}. 
In the simplest case of one electric and one magnetic string, we have\
$(Q_E^1,Q_B^1)=(q_e,0)$ and $(Q_E^2,Q_B^2)=(0,q_m)$, and the moduli
values are\ $(X_1,0)$ and $(0,Y_1)$, so
\be\label{phi-0:modConst}
\tan\varphi = { (0+Y_0)Q_E^1 - (0+X_0)Q_B^2 \over
  (X_1+X_0)Q_E^1 + (Y_1+Y_0)Q_B^2}\ \ra 0 \quad\Rightarrow\quad
{1\over q_e} X_0 = {1\over q_m} Y_0\ .
\ee
This is the moduli constraint (\ref{moduliConstr-111001}) encoding
the asymptotic shape of the M5-brane deformation.
So 
\be\label{T-1string}
T = (X_1+X_0) q_e + (Y_1+Y_0)q_m\ .
\ee
This effective tension can be interpreted in a simple way from the
geometry in Figure~\ref{m5webprongs} (d), as the sum of the tensions
of a 3-pronged M2-brane web comprising
\bea\label{M2-111001-legs}
&&   {\rm leg}_0\,,\ \ {\rm charge} (1,1): \qquad (-X_0,-Y_0) \ra (0,0)\ ,\nn\\
&&   {\rm leg}_1\,,\ \ {\rm charge} (1,0): \qquad (0,0) \ra (X_1,0)\ ,\nn\\
&&   {\rm leg}_2\,,\ \ {\rm charge} (0,1): \qquad (0,0) \ra (0,Y_1)\ ,
\eea
Taking the membrane tension of a $(m,n)$ charge M2-brane as
$\sqrt{m^2q_e^2+n^2q_m^2}$\,, the tension of the composite M2-brane
web above, comprising legs\,$_{0,1,2}$, is
\be\label{M2-111001-tensionSum}
T = \sqrt{q_e^2+q_m^2}\,\sqrt{X_0^2+Y_0^2}\ +\
\sqrt{q_e^2+0}\, X_1\ +\ \sqrt{0+q_m^2}\,Y_1
\ee
which is identical to (\ref{T-1string}) using (\ref{moduliConstr-111001}).
We note that as a 1-string state tension, $T\leq T_1+T_2$, \ie\ $T$
is always less than the 2-string tension
\be\label{T1T2-2string}
T_1+T_2 = q_e\sqrt{(X_1+X_0)^2+(Y_0)^2}\, +\, q_m\sqrt{(X_0)^2+(Y_1+Y_0)^2}\ ,
\ee
which is obtained from the lengths of two independent M2-branes
stretched between the $(X,Y)$-locations\
$\big\{(-X_0,-Y_0),\ (X_1,0)\big\}$\ and\
$\big\{(-X_0,-Y_0),\ (0,Y_1)\big\}$ respectively (see the transverse
geometry in Figure~\ref{m5webprongs} (d)).\
The location in moduli space where the 1-string state becomes equal
in tension to the 2-string state is the wall of marginal stability
(represented by the circle in $X,Y$-space in Figure~\ref{m5webprongs} (d)):
\be\label{X0-0-wms}
X_0, Y_0 \ra 0 :\qquad T \ra T_1+T_2\ .
\ee
This confirms the decay process of the string soliton bound state
into the 2-string state at the wall of marginal stability 
discussed around (\ref{111001-WMSlimit}) using the field
configurations earlier.\ 
It is worth noting that the effective tension (\ref{T-bound}) is
simply a phrase for the mass per unit length of this composite object
near marginal stability where it can be interpreted as approaching the
combined tension of the two constituent strings.

Finally, to recognize the geometry of the locus of marginal stability
as a circle here, it is helpful to define the 2-dimensional vectors in
the $(X,Y)$-plane,
\be\label{xiExiB}
\xi_E \equiv (X_1,0)-(-X_0,-Y_0) = (X_1+X_0,Y_0)\,,\qquad
\xi_B \equiv (0,Y_1)-(-X_0,-Y_0) = (X_0,Y_1+Y_0)\,.
\ee
From Figure~\ref{m5webprongs} (d)), we see that $\xi_E,\ \xi_B$
stretch between the locations
$\{ (-X_0,-Y_0),\ (X_1,0)\}$ and $\{(-X_0,-Y_0),\ (0,Y_1)\}$
respectively.
Then, using the cross-product for 2-vectors, we have
\bea
&&
q_e^2 (\xi_E\cdot\xi_E) + q_m^2 (\xi_B\cdot\xi_B) + 2 q_eq_m |\xi_E\times\xi_B|
\nn\\
&& =\, q_e^2 \left((X_1+X_0)^2+Y_0^2\right) + q_m^2\left(X_0^2+(Y_1+Y_0)^2\right)
+ 2q_eq_m \left((X_1+X_0)(Y_1+Y_0)-X_0Y_0\right) \nn\\
&& =\, (q_e(X_1+X_0) + q_m(Y_1+Y_0))^2 + (q_eY_0-q_mX_0)^2\,.
\eea
This is identical to $T^2$ simplifying $T$ in (\ref{T-bound}) with
the (single electric and single magnetic) charges $(Q_E^1,Q_B^1)=(q_e,0)$
and $(Q_E^2,Q_B^2)=(0,q_m)$ here,
and the corresponding moduli values\ $(X_1,0)$ and $(0,Y_1)$
respectively\footnote{If we use (\ref{moduliConstr-111001}), then the
  last bracket vanishes giving $T$ in (\ref{T-1string}); the angle
  $\varphi$ amounts to an overall $(X,Y)$-rotation ensuring that
  electric/magnetic string spikes stretch along $X/Y$-directions
  respectively.}.
Thus, using $\xi_E,\ \xi_B$ in (\ref{xiExiB}), we see that the
effective 1-string tension $T$ in (\ref{T-bound}) and the 2-string
tension (\ref{T1T2-2string}) can be recast as
\bea\label{T-T1T2-xiExiB}
&& T = \sqrt{q_e^2 (\xi_E\cdot\xi_E) + q_m^2 (\xi_B\cdot\xi_B)
  + 2 q_eq_m |\xi_E\times\xi_B|}\ ,\nn\\ [1mm]
&&\qquad
T_1+T_2 = q_e \sqrt{(\xi_E\cdot\xi_E)} + q_m \sqrt{(\xi_B\cdot\xi_B)}\ .
\eea
Thus\ $T=T_1+T_2$\ when\ $|\xi_E\times\xi_B| = |\xi_E||\xi_B|$,
\ie\ when $\xi_E$ is perpendicular to $\xi_B$, which is equivalent
to the limit (\ref{X0-0-wms}). This describes a
circle in the $(X,Y)$-plane (with $\xi_E, \xi_B$ then subtending an
angle ${\pi\over 2}$ on the circle), which is the locus of marginal
stability shown in Figure~\ref{m5webprongs} (d). The algebraic
manipulations here are of course more or less identical to those
in the D3-brane case (App.~\ref{sec:D3prongs}) with point particle
bound states in
\cite{Argyres:2001pv,Argyres:2000xs,Narayan:2007tx}:\ in particular
see Fig.~4 in \cite{Argyres:2001pv}. This apparent similarity in these
expressions between the M5-brane and the D3-brane cases stems from the
shape of the M5-prong geometry in the transverse space approaching two
separate spikes in the marginal stability limit. In this limit, as we
have seen, the two strings are well-separated with the moduli
variations along the strings ``straightening out''\
(Table~\ref{tableXYbndy21}). Away from this limit, the shape of the
M5-prong configuration is ``blob-like'' and ill-defined in the
transverse $(X,Y)$-space. This also reflects in the M5-brane worldvolume,
where the two strings are not well-separated relative to their
``thickness''. In the analysis above recasting the boundary terms in
terms of an effective 1-string tension approaching the 2-string
tension, it was crucial to write the boundaries in terms of distinct
3-spheres surrounding the strings.  We will discuss this in greater
detail in what follows, obtaining further perspective on this analysis
near marginal stability.

\subsection{More on Bogomolny and the 1-string tension on-shell}
\label{sec:Bog1str-os}

In the previous section~\ref{sec:tension}, we have seen that the
effective 1-string tension obtained upon integrating the energy
functional (\ref{completiongen}) acquires a relatively simple form
near marginal stability, as in eq. (\ref{M2-111001-tensionSum}).  In
particular, note that the energy functional (\ref{completiongen}) is
not written in strict Bogomolny form (perfect squares plus boundary
terms off-shell, such as in eq.(\ref{Bogomolny:D3}) in the D3-brane
case reviewed in App.~\ref{sec:D3prongs}) since there are extra
cross-terms, which however give vanishing contribution on-shell in the
vicinity of marginal stability\ (note that the $4,5$-directions are
noncompact here; compactifying these is tantamount to regarding the
$4,5$-component terms as negligible, which then
effectively reduces (\ref{BogomolnyCompnnts-2}) to (\ref{Bogomolny:D3})
upon integrating over 5-space). We will discuss this in a slightly
more general way now, noting that near marginal stability we have two
well-separated constituent strings that are loosely bound.

The energy functional (\ref{completiongen}) contains two sets of
cross-terms: the second set containing $\zeta^{1} \cdot \zeta^{2}$
vanishes when the constituent strings are orthogonal. 
The first set can be recast as
\begin{align}
& \sum_{a \ne b} \zeta^{1}_a  \zeta^{2}_b \left( \partial_a X \partial_b Y - \partial_b X \partial_a Y\right) = \sum_{a \ne b} \left( \zeta^{1}_a  \zeta^{2}_b \, - \, \zeta^{1}_b  \zeta^{2}_a\right) \partial_a X \partial_b Y \, \cr
&\quad =\, \sum_{a \ne b} \left[ \partial_b \Big( \left( \zeta^{1}_a  \zeta^{2}_b \, - \, \zeta^{1}_b  \zeta^{2}_a\right) (\partial_a X)  Y \Big) \, - \, \left( \zeta^{1}_a  \zeta^{2}_b \, - \, \zeta^{1}_b  \zeta^{2}_a\right) ( \partial_b \partial_a X)  Y \right] .
\end{align}
The second set of terms here vanishes (summing symmetric and antisymmetric
pieces), leaving a total derivative term. Integrating the first set of
terms over 5-space with boundaries being $S^3_\infty$ and the two $S^3$s
enclosing each of the two (well-separated) strings gives
\begin{align} 
& \, \sum_{a \ne b}  \int d^5 x \, \partial_b \Big( \left( \zeta^{1}_a  \zeta^{2}_b \, - \, \zeta^{1}_b  \zeta^{2}_a\right) (\partial_a X)  Y \Big) \cr
   &\quad  = \, \int dl_{\text{string}_1} \sum_{a \ne b} \int \left( d\hat{S}^3 \right)^b \,  \Big(  \left( \zeta^{1}_a  \zeta^{2}_b \, - \, \zeta^{1}_b  \zeta^{2}_a\right) (\partial_a X)  Y \Big) \Big|_{\text{string}_1 } \cr
  & \hspace{1.5cm}  \, + \, \int dl_{\text{string}_2} \sum_{a \ne b} \int \left( d\hat{S}^3 \right)^b \,  \Big(  \left( \zeta^{1}_a  \zeta^{2}_b \, - \, \zeta^{1}_b  \zeta^{2}_a\right) (\partial_a X)  Y \Big) \Big|_{\text{string}_2 }
\end{align}
since the $S^3$ at infinity is at constant modulus value $X=-X_0$ (so
its derivative vanishes). Each string in 5-dimensions is akin to a
point particle in the transverse 4-dimensions: we draw a 3-sphere
$S^3$ surrounding the string. The measure $dl_{\text{string}_i}$ is the
length element along the string $i$ with $i=1,2$, and
$( d\hat{S}^3 )^b$ are the differential 3-volume elements
transverse to the strings, and the $b$-index indicates its orientation
in the 4-dim spatial boundary defined surrounding the respective
strings. The string orientations are defined by the $\zeta$-vectors
which lie in the $(x^4,x^5)$-plane and the $b$-index takes values
accordingly.

Near marginal stability we have $X_0, Y_0$ small: the strings are
well-separated.  We now recast the above boundary terms as
\begin{align} \label{totalderivativecrossT1integrated}
    & \, \sum_{a \ne b}  \int d^5 x \, \partial_b \Big(  \left( \zeta^{1}_a  \zeta^{2}_b \, - \, \zeta^{1}_b  \zeta^{2}_a\right)  (\partial_a X)  Y \Big) \cr
    &\quad = \,   \int dl_{\text{string}_1} \sum_{a \ne b_{\perp_1}} \int \left( d\hat{S}^3 \right)^{b = b_{\perp_1}} \,  \Big(  \left( \zeta^{1}_a  \zeta^{2}_{b_{\perp_1}} \, - \, \zeta^{1}_{b_{\perp_1}}  \zeta^{2}_a\right)  (\partial_a X)  Y \Big) \Big|_{\text{string}_1 } \cr
    & \hspace{1.5cm}  \, + \,   \int dl_{\text{string}_2} \sum_{a \ne b_{\perp_2}} \int \left( d\hat{S}^3 \right)^{b = b_{\perp_2}} \,  \Big(   \left( \zeta^{1}_a  \zeta^{2}_{b_{\perp_2}} \, - \, \zeta^{1}_{b_{\perp_2}}  \zeta^{2}_a\right)  (\partial_a X)  Y \Big) \Big|_{\text{string}_2 } \,
\end{align}
Here $b_{\perp}$ indicates the direction in the $(x^4,x^5)$-plane
orthogonal to the respective strings.

Analysing further, in the string-1 terms (second line above), the
component $\zeta^{1}_{b_{\perp_1}}$ is zero (since $\zeta^1$ has no
component orthogonal to $\text{string}_1$), and further
$\zeta^1_a\del_aX=0$ from translation invariance along string-1. Thus
these string-1 terms in the second line vanish.

Likewise in the string-2 terms (third line above), the 
component $\zeta^{2}_{b_{\perp_2}}$ is zero (since $\zeta^2$ has no
component orthogonal to $\text{string}_2$). The remaining term can be
rewritten as
\begin{align} \label{totalderivativecrossT1integrated2}
     - \,  \sum_{a \ne b_{\perp_2}}  \int \left( d\hat{S}^3 \right)^{b = b_{\perp_2}} \, \int dl_{\text{string}_2}  \Big(   \left( \zeta^{1}_{b_{\perp_2}}  \zeta^{2}_a\right)  (\partial_a X)   \Big) \Big|_{\text{string}_2 } \times Y_1 \,.
\end{align}
(recall that $Y_1$ is a constant cut-off value for the scalar $Y$
near the location of $\text{string}_2$)
Now we note that $X$ is an even function along string-2 (symmetric
about the point of closest approach of string-2 with string-1). Thus
its derivative $\zeta^2_a\del_aX$ is an odd function along string-2:
so integrating along the entire length of string-2 gives a vanishing
contribution. Thus this term also vanishes.

Let us consider the simplest case of two orthogonal strings to
illustrate how the above arguments work.
To compare with our analysis in sec.~\ref{sec:tension}, recall 
that (from (\ref{m5m2-111001}))\ $\del_4X=0$,\ and\
$X = \frac q{ |\vec{r} - \vec{r}_1|^2 +  (x^5)^2} - X_0$.\ Thus 
the first set of terms gives\
\begin{align}\label{totalderivativecrossT1integratedOrthoX&Y}
        \int dl_{\text{string}_1} \int \left( d\hat{S}^3 \right)^{ b_{\perp_1} = 5} \,  \Big(  \left( \zeta^{1}_4  \zeta^{2}_{5} \right)  (\partial_4 X)  Y \Big) \Big|_{\text{string}_1 }  
\end{align}
which thus vanishes. Note that the $S^3$ here surrounds string$_1$
embedded in the $\BR^4$ orthogonal to the $x^4$-direction, so that only
$b_{\perp_1} = 5$ appears in the $S^3$-element above. Likewise, the
terms (\ref{totalderivativecrossT1integrated2}) here become
\begin{align} \label{totalderivativecrossT1integarted2OrthoX&Y}
     - \,  \int \left( d\hat{S}^3 \right)^{ b_{\perp_2} =4} \, \int dx^5 \, \Big( \left( \zeta^{1}_{b_{\perp_2} =4} \,  \zeta^{2}_5 \right)  (\partial_5 X)   \Big) \Big|_{\text{string}_2 } \times Y_1 \,,
\end{align}
and $\del_5X$ is odd, so the integral vanishes.


Thus overall, we find these extra cross-terms in the Bogomolny
completion vanish on-shell, when integrated over the 5-space with
appropriate boundaries around the two well-separated strings near
marginal stability (when the constituent strings are not orthogonal
there are further terms as well, the $\zeta^{1} \cdot \zeta^{2}$ terms
in (\ref{completiongen})). Of course, in effect we have simply recast
the earlier analysis leading to (\ref{M2-111001-tensionSum}) in a
slightly more general manner, but this is perhaps useful in
illustrating these extra terms as vanishing boundary terms on-shell.
The essential ingredients here of course is that the strings are
well-separated so one can break up the 5-space integral into boundary
terms over appropriate separate $S^3$s surrounding the two
strings. This is of course justified near marginal stability and
becomes increasingly accurate here, and is consistent with the fact
that our entire analysis is only reliable near marginal
stability. However it begs the question of finding ``strict Bogomolny
completions'' that are manifestly of the form ``perfect-squares plus
boundary terms'' but at an \emph{off-shell} level.

From the point of view of regarding one string as background and the
other as a probe, we expect that the probe string will feel larger
3-form and scalar forces near its ``bulk'' which is closer to the bulk
of the background string. This suggests that it will bend resulting in
curved string profiles, unlike the straight string configurations
(Figure~\ref{m5webprongs} (c)) we have been studying. These bent (or
curved) string equilibrium configurations will likely be solutions to
different BPS equations following from some possibly different
Bogomolny energy extremization, holding different boundary terms
fixed, with no extra cross-terms (looking for such bent string
configurations appears difficult however!). Near marginal stability,
the strings are well-separated and so will straighten out (with
negligible bending), thus coinciding with our picture of straight
non-parallel loosely bound strings near marginal stability.  In this
light, it is perhaps not surprising that these cross-terms above only
vanish on-shell.

\section{More general M5-M2 prong states}\label{sec:m5m2gen}

We now consider more general M5-M2 prong bound states: in general
these have charges $(p_1+p_2,q_1+q_2)-(p_1,q_1)-(p_2,q_2)$ and
corresponding nontrivial angles between the constituent legs. The
field configurations can be constructed using two general
$\zeta$-vectors with $X',Y'$ defined appropriately as in
(\ref{genBPS-X'Y'}). We will however find it more instructive to
discuss these illustrating them with two concrete nontrivial examples.

\subsection{$(2,1)-(1,0)-(1,1)$}\label{subsection(2,1)}

Consider the charges $(2,1)-(1,0)-(1,1)$. We will keep the factors
$q_e, q_m$ explicit.  This configuration is captured by the
$\zeta$-vectors and the string parametrizations via $(x^4,x^5)$-coordinates,
\vspace{-3mm}
\begin{subequations}\label{211011-zetaStringCoords}
	\begin{align}
	\label{211011-zetaStringCoordsa}
	\zeta_1 = (1,0) : & \qquad \,\,
	\vx_1\equiv (x^4,x^5)_1 = x_1 (1,0)\quad \forall\ x_1\,,\quad \vr=\vr_1\,,\\
	\zeta_2 = {1\over\sqrt{1+{q_e^2\over q_m^2}}}\Big({q_e\over q_m}\,,1\Big)
	\equiv \gamma (\beta ,1) : &  \qquad \,\,
	\vx_2\equiv (x^4,x^5)_2 = x_2 (\beta ,1)\quad \forall\ x_2\,,\quad
	\vr=\vr_2\,. \,
	\label{211011-zetaStringCoordsb}
	\end{align}
\end{subequations}
We have used the translation symmetry along the $x^4$ and $x^5$
directions to parametrize the string coordinates so that they always
pass through the origin $(x^4,x^5)=(0,0)$. Projecting these lines onto
the $(x^4,x^5)$-plane, these lines can always be seen to intersect at
the origin which is their closest point of approach in this convenient
parametrization.

Using the ``basis'' field configurations (\ref{XYbasisConfig}) with\
$c_Y\equiv q_m/\gamma$\ and\ $c_X\equiv q_e$\,,\
the BPS equations (\ref{genBPS-X'Y'}) give the solutions
\bea\label{m5m2-211011-'}
&& X' = {q_e\over (\vr-\vr_1)^2 + (\zeta_1^\perp\cdot(\vx-\vx_1))^2}
+ {q_e\over (\vr-\vr_2)^2 + (\zeta_2^\perp\cdot(\vx-\vx_2))^2} - X_0'\ ,
\nn\\
&& Y' = {q_m\over (\vr-\vr_2)^2 + (\zeta_2^\perp\cdot(\vx-\vx_2))^2} - Y_0'\ .
\eea
The denominators here encode the transverse distance from any point on
string$_1$ or string$_2$: since we have translation invariance along
the strings, this means the ``equipotential surfaces'' for string$_1$
are concentric cylinders around it, and likewise for string$_2$.
Thus if we translate along the string, the corresponding field
configuration must remain constant so the location along the string
should scale away from these expressions: this can be seen explicitly
by noting that $\zeta_1^\perp\cdot \vx_1=0$ and
$\zeta_2^\perp\cdot \vx_2=0$ using the parametrizations
(\ref{211011-zetaStringCoords}). This gives
\bea\label{m5m2-211011}
&& X' = {q_e\over (\vr-\vr_1)^2 + (\zeta_1^\perp\cdot\vx)^2}
+ {q_e\over (\vr-\vr_2)^2 + (\zeta_2^\perp\cdot\vx)^2} - X_0'\ ,
\nn\\
&& Y' = {q_m\over (\vr-\vr_2)^2 + (\zeta_2^\perp\cdot\vx)^2} - Y_0'\ .
\eea
These then give the electric and magnetic components $\vE, \vB$ of
the field strength. As $\vr\ra\infty$ we have 
\be
\vr\ra\infty,\ \ x^4, x^5\ {\rm finite}:\qquad
X'\sim {2q_e\over |\vr|^2}-X_0'\ ,\qquad Y'\sim {q_m\over |\vr|^2}-Y_0'\ ,
\ee
so we find $(\vE,\vB)=(\nabla X',\nabla Y')\sim (2,1)$ encoding the
asymptotic charge of this configuration. Thus this M5-brane
deformation in this asymptotic regime traces a line along $(2,1)$ in
the $(X',Y')$ moduli space transverse to the M5-brane when
\be\label{modulishape21}
{1\over q_e}\,X_0' = {2\over q_m}\,Y_0'\ .
\ee

Now we study the scalar field configurations as we approach the
strings. First, analogous to (\ref{bc-111001}), let us impose the
boundary conditions (\ref{bc-gen}). Then from the $Y'$ configuration,
we have $Y'\ra 0$ which gives
\be\label{211011-M5wms}
{1\over (\vr_1-\vr_2)^2} = {1\over q_m} Y_0' = {1\over 2q_e} X_0'\ .
\ee
This holds at $(x^4,x^5)=(0,0)$. Away from this, as we go along
string$_1$ stretched along $\vrho=(\vr_1,x_1,0)$, the $Y'$ scalar
behaves as
\be
Y' \sim {q_m\over (\vr_1-\vr_2)^2 + \gamma^2 x_1^2} - Y_0'\quad
\xrightarrow{\ |x_1|\ra\infty\ }\quad  - Y_0'\ ,
\ee
using $\zeta_2^\perp=\gamma (1,-\beta)$.\ 
Likewise one can analyse the $X', Y'$ field configurations in
various asymptotic limits, as we go along string$_1$ or string$_2$.
This is most efficiently done using the relations (\ref{genBPS-X'Y'})
in terms of the ``basis fields'' $X, Y$:\ using
(\ref{211011-zetaStringCoords}) these become
\be\label{XYvsXpYp21}
X' = X + \beta \gamma \, Y \ ,\qquad Y' = \gamma \, Y \,,
\ee
so
\be
Y'_0 = \gamma Y_0\ ,\quad X'_0 = X_0 + \beta \gamma \, Y_0\ ,
\quad X'_1 = X_1\ ,\quad Y'_1 = \gamma Y_1\ .
\ee
The resulting scalar asymptotics on the moduli space for $X', Y'$,
in various limits can be written by using those for $X, Y$ in
Appendix~\ref{App:details2132} and are tabulated in
Table~\ref{tableXYprimebndy}.

\begin{table}[H]
	\begin{center}
		\begin{tabular}{|c|c|c|}
			\hline
			$(X', Y') \rightarrow (X'_1 \, , 0)$ &  $\quad\text{}$ $\vr \ra \vr_1 \quad$& $x_1=0$ \cr
			\hline
			$(X', Y') \rightarrow (X'_1 - \beta \, Y'_0 \,\,, -Y'_0)$ &  $\vr \ra \vr_1$ & $|x_1| \ra \infty$  \cr
			\hline
			$(X', Y') \rightarrow (\beta  \, Y'_1 \,\, ,  \, Y'_1)$ &  $\vr\ra \vr_2$ & $x_2=0$ \cr
			\hline
			$(X', Y') \rightarrow (- \left( X'_0 - \beta  Y'_0 \right) + \beta Y'_1  \,,\, Y'_1)$ &  $\vr\ra \vr_2$ & $|x_2|\ra\infty$ \cr
			\hline
			$(X',Y') \rightarrow (-X'_0 \,, - Y'_0)$ & $\vr\ra \infty$ & $|x_1|$, $|x_2|$ $\rightarrow \infty$  \cr
			\hline
		\end{tabular}
	\end{center}
	\caption{$X', Y'$ moduli asymptotics} 
	\label{tableXYprimebndy}
\end{table}
It is clear that in the limit of approaching the wall of marginal
stability
\be
X_0',\ Y_0'\ \ll\ X_1',\ Y_1'\ ,
\ee
the M5-brane bendings ``straighten out'' and the moduli approach
(i) $(X_1',0)$ along the entire string$_1$, \ie\ as
$\vrho\ra (\vr_1,x_1,0)$,\ and\ (ii) $(\beta Y_1', Y_1')$
along the entire string$_2$, \ie\ as $\vrho\ra (\vr_2,\beta x_2,x_2)$.
This is similar to the results in the previous section where we
discussed the $(1,1)-(1,0)-(0,1)$ M5-brane prongs.

\subsection{$(3,2)-(1,1)-(2,1)$}\label{subsection(3,2)}

The analysis for this example is similar to the previous case but is a
bit more involved.  This configuration has the $\zeta$-vectors and the
coordinate parametrizations on the strings,
\begin{subequations}\label{321121-zetaStringCoords}
	\begin{align}
	\label{321121-zetaStringCoordsa}
	\zeta_{1} = {1\over\sqrt{1+{q_m^2\over q_e^2}}}\Big(1 \, ,{q_m\over q_e}\Big) \equiv \gamma_1 (1 \, ,\beta_1)  :\  & \qquad
	\vx_1\equiv (x^4,x^5)_1 = x_1 (1 \, ,\beta_1 ) \quad \forall\ x_1\,,\quad \vr=\vr_1\,,\\
	\zeta_{2} = {1\over\sqrt{1+{( 2 q_e)^2\over q_m^2}}}\Big({2 q_e\over q_m}\,,1\Big)
	\equiv \gamma_2 (\beta_2\,,1)  :\  &  \qquad
	\vx_2\equiv (x^4,x^5)_2 = x_2 (\beta_2 ,1)\quad \forall\ x_2\,,\quad
	\vr=\vr_2\,. \qquad 
	\label{321121-zetaStringCoordsb}
	\end{align}
\end{subequations}
Using the field configurations in (\ref{XYbasisConfig}) with \ 
$c_X\equiv q_e / \gamma_1$\ and\ $c_Y\equiv q_m/\gamma_2$ \,,\
the BPS equations (\ref{genBPS-X'Y'}) give the solutions
\bea\label{m5m2-321121}
&& X' = {q_e\over (\vr-\vr_1)^2 + (\zeta_1^\perp\cdot\vx)^2}
+ {2 q_e\over (\vr-\vr_2)^2 + (\zeta_2^\perp\cdot\vx)^2} - X_0'\ ,
\nn\\
&& Y' = {q_m\over (\vr-\vr_1)^2 + (\zeta_1^\perp\cdot\vx)^2}
+ {q_m\over (\vr-\vr_2)^2 + (\zeta_2^\perp\cdot\vx)^2} - Y_0'\ ,
\eea
which also give the $\vE, \vB$ components of the field strength tensor.
As $\vr\ra\infty$ we see that
\be\label{321121-rInfty}
\vr\ra\infty,\ \ x^4, x^5\ {\rm finite}\,:\qquad
X'\sim {3q_e\over |\vr|^2}-X_0'\ ,\quad Y'\sim {2q_m\over |\vr|^2}-Y_0'\ ,
\ee
and the electric and magnetic field components scale as
$(\vE,\vB)\propto (3,2)$, thereby encoding charge $(3,2)$.
The M5-brane deformation in this asymptotic regime traces a line
along $(3,2)$ in the $(X',Y')$ moduli space when
\be\label{modulishape32}
{2\over q_e}\,X_0' = {3\over q_m}\,Y_0'\ .
\ee
We impose the boundary conditions
\be
\vrho\ra (\vr_1,0,0):\ \ 
(X',Y')\ra \gamma_1 \Big( X_1 , { q_m \over q_e } X_1  \Big)\,;\quad\
\vrho\ra (\vr_2,0,0):\ \
(X',Y')\ra \gamma_2 \Big(Y_1{2 q_e\over q_m}\,,Y_1\Big)\ .
\ee
This fixes the transverse distance between the strings in terms of
the scalar moduli values
\be\label{321121-M5wms}
{1\over (\vr_1-\vr_2)^2} = {1\over 2 q_m} Y_0' = {1\over 3 q_e} X_0'\ .
\ee

\noindent
As in the previous example, we can analyse the $X', Y'$ field
configurations in various asymptotic limits, as we go along
string$_1$ or string$_2$.
This is most efficiently done using the relations (\ref{genBPS-X'Y'})
in terms of the ``basis fields'' $X, Y$:\ using
(\ref{321121-zetaStringCoords}) these become
\be\label{XYvsXpYp32}
X' = \gamma_1 X + \beta_2 \gamma_2 \, Y \ ,\qquad Y' = \gamma_1 \beta_1 X +  \gamma_2 \, Y \,,
\ee
so
\begin{align}
X'_0 = \gamma_1 X_0 + \beta_2 \gamma_2 \, Y_0 \ ,&\qquad Y'_0 = \gamma_1 \beta_1 X_0 +  \gamma_2 \, Y_0  ,\cr 
\qquad X'_1 = \gamma_1 X_1 \,\,,& \quad Y'_1 = \gamma_2 Y_1
\end{align}
The resulting scalar asymptotics on the moduli space for $X', Y'$ are
tabulated in Table~\ref{tableXYprimebndy32} (see
Appendix \ref{App:details2132}).

\begin{table}[H]
	\begin{center}
		\begin{tabular}{|c|c|c|}
			\hline
			$(X', Y') \rightarrow ( X'_1 \, ,  \beta_1 \, X'_1)$ &  $\quad\text{}$ $\vr \ra \vr_1 \quad$& $x_1=0$ \cr
			\hline
			$(X', Y') \rightarrow \left( \, X'_1 - \left( 2 X'_0 - \beta_2 Y'_0 \right) \,\,, \,\, \beta_1 X'_1 - \left( \beta_1 X'_0 -  Y'_0 \right)\, \right)$ &  $\vr \ra \vr_1$ & $|x_1| \ra \infty$  \cr
			\hline
			$(X', Y') \rightarrow ( \beta_2  \, Y'_1 \,\, ,  \, Y'_1)$ &  $\vr\ra \vr_2$ & $x_2=0$ \cr
			\hline
			$(X', Y') \rightarrow \left( \,  \beta_2 Y'_1 - \left( \beta_2 Y'_0 -  X'_0 \right)   \,,\, Y'_1 - \left( 2 Y'_0  - \beta_1 X'_0 \right) \, \right)$ &  $\vr\ra \vr_2$ & $|x_2|\ra\infty$ \cr
			\hline
			$(X',Y') \rightarrow (-X'_0 \,, - Y'_0)$ & $\vr\ra \infty$ & $|x_1|$, $|x_2|$ $\rightarrow \infty$  \cr
			\hline
		\end{tabular}
	\end{center}
	\caption{$X', Y'$ moduli asymptotics for $(3,2) - (1,1) - (2,1)$} 
	\label{tableXYprimebndy32}
\end{table}
\noindent
As we approach the wall of marginal stability,
\be
X_0',\ Y_0'\ \ll\ X_1',\ Y_1'\ ,
\ee
the moduli approach
(i) $(X_1',\beta_1 X'_1)$ along the entire string$_1$, \ie\ as
$\vrho\ra (\vr_1,  x_1, \beta_1 x_1)$,\ and\ (ii) $(\beta_2 Y_1', Y_1')$
along the entire string$_2$, \ie\ as $\vrho\ra (\vr_2,\beta_2 x_2,x_2)$.
This is similar to the previous cases,
$(1,1)-(1,0)-(0,1)$ and $(2,1)-(1,0)-(1,1)$.

\subsection{1-string tension}\label{tensionnon-ortho}

Along the same lines as in Sec.~\ref{sec:tension}, we can calculate
the tension of the M2-brane web for the configurations we have described.
For the $(2,1)-(1,0)-(1,1)$ M2-web, we obtain
\be\label{mass-tension(2,1)}
T_{(2,1)} \,=\,  L \left(  \, \left(  X'_1 \, + \, \beta \, Y'_1  \,
+ \, 2 X'_0  \right) q_e \, + \,  \left( Y'_1 + Y'_0 \right) q_m \, \right) ,
\ee
while for the $(3,2)-(1,1)-(2,1)$ M2-web, we obtain
\be\label{mass-tension(3,2)}
T_{(3,2)} \,=\, L \left(  \, \left(  X'_1 \, + \, 2 \beta_2 \, Y'_1  \,
+ \, 3 X'_0  \right) q_e \, + \,  \left(  \beta_1 \, X'_1  + Y'_1 + 2 Y'_0
\right) q_m \, \right) .
\ee
These can again be interpreted in terms of the tensions of three
M2-brane legs of appropriate charge and shapes, along the lines of
(\ref{M2-111001-legs}) and (\ref{M2-111001-tensionSum}), using the
$(m,n)$ membrane tension and the moduli constraints in
(\ref{modulishape21}) and (\ref{modulishape32}) respectively.
Various details are described in App.~\ref{App:details2132} and
specifically App.~\ref{sec:App-tension2132}. 

In the formula for energy functional (\ref{BogomolnyCompnnts}) there
is another term which could give contribution to the tensions in
(\ref{mass-tension(2,1)}) and (\ref{mass-tension(3,2)}). The cross
term: $\left( \zeta^1 \cdot \zeta^2 \right)\left( \partial X \cdot
\partial Y \right) $ in (\ref{BogomolnyCompnnts}) is non-zero for the
examples above with two non-orthogonal strings. In
App.~\ref{sec:App-tension2132}, we analyze this term in some detail
and discuss the non-trivial contribution that could come from this
term. However, in the limit where we consider the transverse distance
between the line sources to be large, we find that the contribution
from this cross term becomes negligible. Thus near the wall of
marginal stability limit (where $X_0, Y_0 \sim 0$) the collective
tensions of the configuration in these examples essentially approach
those in (\ref{mass-tension(2,1)}) and (\ref{mass-tension(3,2)}),
respectively.

\section{Embedding in ``Higgsed'' M5-brane theories}\label{M5s-higgsed}

We now consider embedding our description of M5-brane prongs
into the theory of three M5-branes, parallel and ``Higgsed'' (\ie\ on
the tensor branch). The schematics with multiple M5s here is similar
to that with multiple D3-branes in \cite{Argyres:2001pv,Narayan:2007tx}.
The energy functional is
\be\label{EngyFnal-3M5s}
\mathcal{E} = \frac12 \sum_{i=1,2,3} \left( |\partial X^{(i)}|^2
+ |\partial Y^{(i)}|^2
+ \frac{1}2|\tilde{\cal H}^{(i)}|^2 \right) ,\qquad
\ee
where each M5--brane\,$_i$ is regarded as independent of the other two.
This is essentially three copies of (\ref{freelimEngy}).
This is reasonable in the low energy abelian approximation where
the only interaction between the branes arises at the locations
where the M2-brane spike legs from one M5-brane join those
from another. These locations necessarily include nonabelian
contributions, lying outside our approximations: we incorporate their
effects through the moduli boundary conditions.

Since the branes are decoupled, we carry out Bogomolny completions
(\ref{completiongen}) independently within each brane$_i$ sector,
giving (\ref{genBogo}) for each sector. Componentwise
these lead via (\ref{BPScomp}) to (\ref{genBPS-X'Y'}) for each
brane$_i$. Since the M2-brane stretching out from one M5-brane
ends on another M5-brane, the line source conditions (\ref{lineSrcXY})
for these must be correlated: likewise the boundary conditions
(\ref{bc-gen}) are correlated. It is easiest to illustrate this
via simple examples.

A simple electrically charged configuration of a single M2-brane
stretching between two M5-branes is described by juxtaposing two
configurations (\ref{M5spike}). Using the form (\ref{M5spike-zeta}),
we have
\bea\label{M5spikeM5}
X^{(1)} = {q_e\over (\vr-\vr_0^{(1)})^2 + (\zeta^\perp\cdot(\vx-\vx_0^{(1)}))^2}
- X_0^{(1)}\,;\ &&\ \vE^{(1)}=\nabla X^{(1)}\,,\quad \Pi^{(1)}=-\del_5X^{(1)}\,,
\nn\\
X^{(2)} = -{q_e\over (\vr-\vr_0^{(2)})^2 + (\zeta^\perp\cdot(\vx-\vx_0^{(2)}))^2}
- X_0^{(2)}\,;\ &&\ \vE^{(2)}=\nabla X^{(2)}\,,\quad \Pi^{(2)}=-\del_5X^{(2)}\,.
\quad\ \
\eea
In order that the M5-spike$_1$ joins with M5-spike$_2$, we must
have
\be
\vr_0^{(1)} = \vr_0^{(2)}\ ,\qquad \vx_0^{(1)} = \vx_0^{(2)}\ .
\ee
The fact that $\vE^{(1)}$ and $\vE^{(2)}$ are opposite in sign is
tantamount to the statement that the M2-brane stretching between the
two separated M5-branes describes a charge $(1,0)_1-(-1,0)_2$ from the
point of view of the $U(1)_1\times U(1)_2$ theory, ``Higgsed'' from
the $U(2)$ theory of two M5-branes.

Now in addition, the fact that the two spike ``cores'' join
at the same location in moduli space requires that they be the same
``core size'' within this abelian approximation: using (\ref{al00-X}),
this gives
\bea
q_e \al_{00}^{(1)} - X_0^{(1)} \sim X_1^{(1)} = X_1^{(2)} \sim
-q_e \al_{00}^{(2)} - X_0^{(2)}\ ,\qquad \quad \nn\\ [1mm]
\al_{00}^{(1)}=\al_{00}^{(2)}\qquad
\Rightarrow\qquad X_0^{(1)} + X_1^{(1)} = - X_1^{(1)} - X_0^{(2)}\ ,\quad
\nn\\ [1mm]
\Rightarrow\qquad X_1^{(1)} = -{1\over 2} (X_0^{(1)} + X_0^{(2)})\ ,\qquad
\al_{00}^{(1)}=\al_{00}^{(2)} = {1\over 2q_e} (X_0^{(1)} - X_0^{(2)})\ .\quad
\eea
The core size thus scales as the inverse distance between the
M5-branes which are located at $-X_0^{(1)}$ and $-X_0^{(2)}$\ (where
the deformations in (\ref{M5spikeM5}) vanish).  Also, we see that the
gluing location in moduli space is at the midpoint between the two
M5-branes. This is the location of enhanced $U(2)$ symmetry in moduli
space where new light nonabelian tensor modes must be added to
the low energy theory for a nonsingular description.

\begin{figure}[h]
\hspace{-0.5pc}
  \includegraphics[width=18pc]{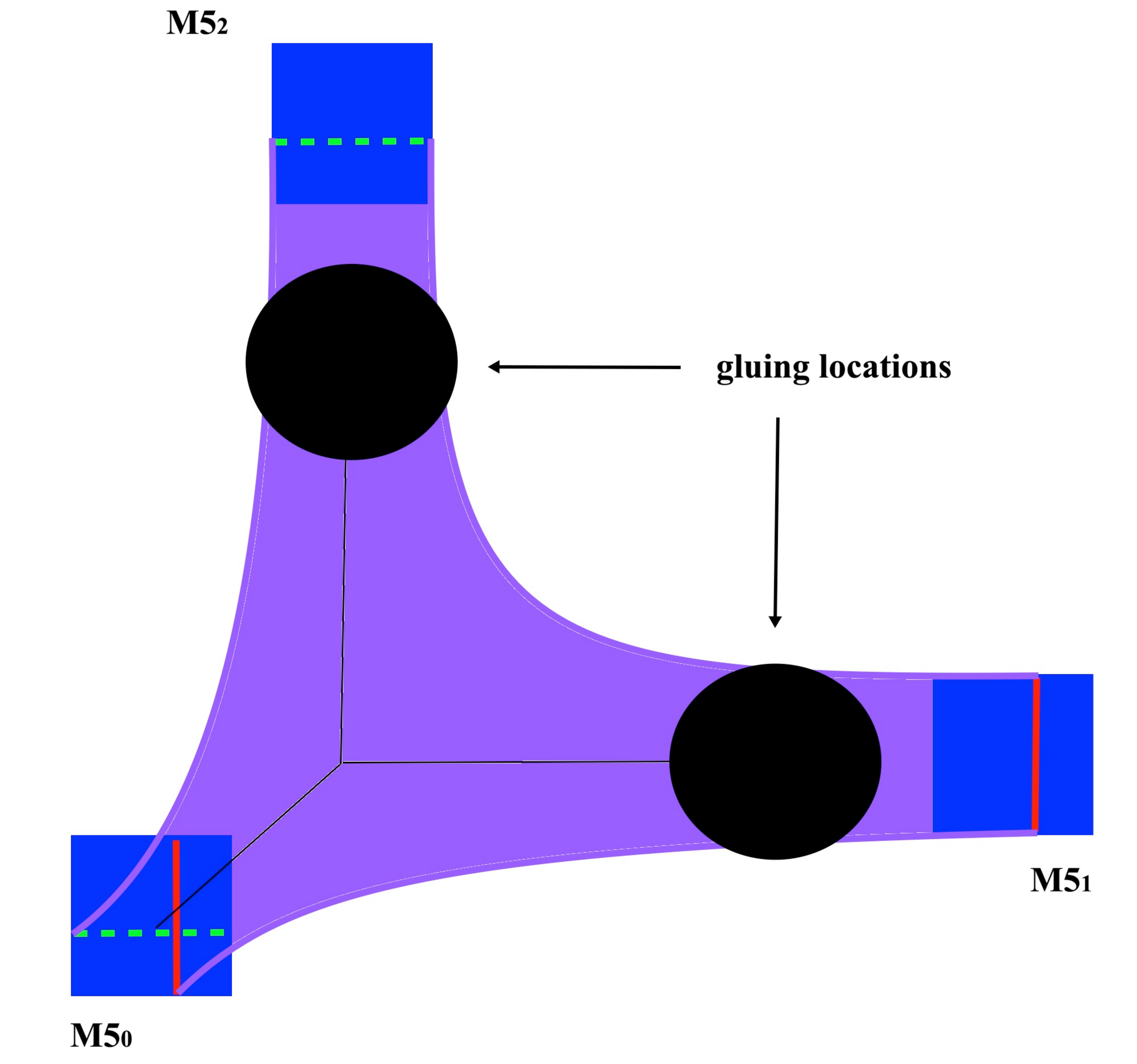}
\hspace{0.75pc}
\begin{minipage}[b]{19.5pc}
  \caption{{ \label{3M5web}
    \footnotesize Cartoon of three parallel separated M5-branes with a
    prong deformation stretching between them in the transverse
    $(X,Y)$-space. The solid red and dashed green lines depict the
    non-parallel M2-endlines in the M5-brane worldvolumes (spanning
    $\{0,1,2,3,4,5\}$). The solid black circles represent the gluing
    locations, where the prong legs are joined to the corresponding
    spike legs. \newline \newline
}}
\end{minipage}
\end{figure}
Along similar lines, we can embed the simplest M5-brane prong
configurations in a theory of three parallel separated M5-branes located
as in Figure~\ref{3M5web} (drawing this accurately requires more
dimensions than possible(!); the directions are as in (\ref{M5M2geom})),
which completes Fig.~\ref{m5webprongs} (d):
\bea\label{M5M5M5-locations}
&& M5\,_0:\ \ (-X_0^{(0)},-Y_0^{(0)}) ;\qquad\ \
M5\,_1:\ \ (X_0^{(1)},0) ;\qquad\ \
M5\,_2:\ \ (0,Y_0^{(2)}) ;\nn\\ [1mm]
&& \qquad\qquad\qquad\qquad\
X_0^{(0)}, Y_0^{(0)} \ll X_0^{(1)} ,\ Y_0^{(2)}\ .
\eea
The last condition encodes the fact that M5-brane\,$_0$ is near the
wall of marginal stability so its $X, Y$ scalars are turned on in a
prong-like field configuration (\ref{m5m2-111001}).
Each constituent spike leg of this prong is glued onto a corresponding
spike (\ref{M5spike}) from one of the other M5-branes. The field
configurations then are\ (with $Y^{(1)}=0$ and $X^{(2)}=0$)
\bea\label{M5prongM5M5}
&& \vE^{(0)}=\nabla X^{(0)}\ ,\qquad
X^{(0)} = {q_e\over (\vr-\vr_1^{(0)})^2 + (\zeta_1^\perp\cdot(\vx-\vx_1^{(0)}))^2}
- X_0^{(0)}\ ,\nn\\
&& \vB^{(0)}=\nabla Y^{(0)}\ ,\qquad
Y^{(0)} = {q_m\over (\vr-\vr_2^{(0)})^2 + (\zeta_2^\perp\cdot(\vx-\vx_2^{(0)}))^2}
- Y_0^{(0)}\ , \nn\\ [1mm]
&& \Pi^{(0)} = \del_4Y^{(0)}-\del_5X^{(0)}\ ,\\ [1mm]
&& \vE^{(1)}=\nabla X^{(1)}\ ,\quad
X^{(1)} = -{q_e\over (\vr-\vr_0^{(1)})^2 + (\zeta_1^\perp\cdot(\vx-\vx_0^{(1)}))^2}
+ X_0^{(1)}\ , \qquad \Pi^{(1)}=-\del_5X^{(1)}\ , \nn\\
&& \vB^{(2)}=\nabla Y^{(2)}\ , \quad
Y^{(2)} = -{q_m\over (\vr-\vr_0^{(2)})^2 + (\zeta_2^\perp\cdot(\vx-\vx_0^{(2)}))^2}
+ Y_0^{(2)}\ , \qquad \Pi^{(2)}=\del_4Y^{(2)}\ .\nn
\eea
The fact that the $\vE^{(0)}, \vE^{(1)}$ are of opposite sign (and
likewise $\vB^{(0)}, \vB^{(2)}$) indicates the fact that the M2-branes
stretching between the three parallel separated M5-branes describes a
charge $(1,1)_0-(-1,0)_1-(0,-1)_2$ configuration from the point of view
of the $U(1)_0\times U(1)_1\times U(1)_2$ theory, ``Higgsed'' from the
$U(3)$ theory of three M5-branes.

In order that the prong leg-$X^{(0)}$ joins with spike$_1$ and
correspondingly prong leg-$Y^{(0)}$ joins with spike$_2$, we must
have
\be
\vr_1^{(0)} = \vr_0^{(1)}\ ,\quad \vx_1^{(0)} = \vx_0^{(1)}\ ;\qquad\quad
\vr_2^{(0)} = \vr_0^{(2)}\ ,\quad \vx_2^{(0)} = \vx_0^{(2)}\ ,
\ee
\ie\ the string locations in the various M5-branes must match, or
equivalently the M2-brane stretching between M5-brane\,$_0$ and
M5-brane\,$_1$ (and likewise M5-brane\,$_0$ and M5-brane\,$_2$)
creates endlines whose worldvolume locations must match.

In addition, the string ``core sizes'' must match between
M5-brane\,$_0$ and M5-brane\,$_1$ (and likewise M5-brane\,$_0$ and
M5-brane\,$_2$). Applying (\ref{111001-coreSize}) and (\ref{al00-X})
to the M5-branes appropriately, this matching requires
\bea
&& q_e \al_{11}^{(0)} - X_0^{(0)} \sim X_1^{(0)}\ =\ X_1^{(1)} \sim 
-q_e \al_{00}^{(1)} + X_0^{(1)}\ ,\qquad \al_{11}^{(0)}=\al_{00}^{(1)}\ ;
\quad \nn\\ [1mm]
&& q_m \al_{22}^{(0)} - Y_0^{(0)} \sim Y_1^{(0)}\ =\ Y_1^{(2)} \sim 
-q_m \al_{00}^{(2)} + Y_0^{(2)}\ , \qquad \al_{22}^{(0)}=\al_{00}^{(2)}\ .
\eea
Solving for the moduli locations $(X_1^{(0)},0)$ and $(0,Y_1^{(0)})$
at which we perform the gluing and the inverse core sizes $\al_{ii}$,
we obtain
\bea
&& \quad
X_0^{(0)} + X_1^{(0)} = - X_1^{(0)} + X_0^{(1)}\ ;\qquad
Y_0^{(0)} + Y_1^{(0)} = - Y_1^{(0)} + Y_0^{(2)}\ , \nn\\ [1mm]
\Rightarrow && \quad
X_1^{(0)} = {1\over 2} ( X_0^{(1)} - X_0^{(0)} )\ ,\qquad
\al_{11}^{(0)}=\al_{00}^{(1)} = {1\over 2q_e} ( X_0^{(0)} + X_0^{(1)} )\ ,
\nn\\ [1mm]
\Rightarrow && \quad
Y_1^{(0)} = {1\over 2} ( Y_0^{(2)} - Y_0^{(0)} )\ ,\qquad
\al_{22}^{(0)}=\al_{00}^{(2)} = {1\over 2q_m} ( Y_0^{(0)} + Y_0^{(2)} )\ .
\eea
Using the locations (\ref{M5M5M5-locations}) of the three M5-branes,
we see that the gluing occurs at the midpoint between M5-brane\,$_0$
and M5-brane\,$_1$ (and likewise M5-brane\,$_0$ and M5-brane\,$_2$).
These are the locations of enhanced gauge symmetry where the stretched
M2-branes become massless signalling light nonabelian degrees of
freedom. The inverse core sizes scale as the distance between the
corresponding pairs of M5-branes. Finally, analysing the moduli values
in the vicinity of the string soliton cores as in \eg\
(\ref{111001-X:x5=0}) suggests the boundary conditions 
\bea\label{bc3M5-111001}
\vrho\ra (\vr_1^{(0)},0,0):\ &&\ (X^{(0)},Y^{(0)}) \ra\ (X_1^{(0)},0)\ \leftarrow
(X^{(1)},Y^{(1)})\ ,\nn\\ 
\vrho\ra (\vr_2^{(0)},0,0):\ &&\ (X^{(0)},Y^{(0)}) \ra\ (0,Y_1^{(0)})\ \leftarrow
(X^{(2)},Y^{(2)})\ ,
\eea
analogous to (\ref{bc-111001}). This leads
to the transverse separation between the string solitons on M5\,$_0$
as\ $(\vr_1^{(0)}-\vr_2^{(0)})^2={q_e\over X_0^{(0)}}$\,, analogous to
(\ref{111001-M5wms}), with a similar wall-crossing limit $X_0^{(0)}\ra 0$.

\section{Discussion}\label{sec:Disc}

We have constructed and described field configurations representing
self-dual string soliton bound states in the M5-brane abelian
effective field theory.  These resemble M5-brane ``prongs'' as
we have seen: the simplest such configuration is depicted in
Figure~\ref{m5webprongs} (to the extent possible: a more full picture
requires more dimensions than can be drawn!).  While the Bogomolny
completion (\ref{completiongen}) and the resulting BPS bound equations
(\ref{genBogo}) appear general and written in terms of intrinsically
M5-brane structures, the explicit field configurations are best
expressed and understood in terms of the component forms
(\ref{BPScomp}), (\ref{genBPS-X'Y'}) and (\ref{BogomolnyCompnnts}),
which decompose the self-dual 3-form tensor field strength
(\ref{def-tildeH}) into appropriate electric and magnetic components
(\ref{HtildeH-comps}). It may be interesting to understand how our
starting point, the effective low energy functional (\ref{freelimEngy}),
gels with the recent action formulations of self-dual forms
\cite{Sen:2015nph,Sen:2019qit} (see also \cite{Andriolo:2020ykk}).

A single string soliton representing the endline of an M2-brane ending
on the M5-brane is described by an M5-brane spike deformation
(\ref{M5spike}) sourced by a single scalar field: an electric string
has field strength components\ ${\tilde H}^{a4},\ {\tilde
  H}^{45}$\ ($a=1,2,3$).\ Bound states of two string solitons
described by M5-brane prong deformations, have nonvanishing field
strength components ${\tilde H}^{a4},\ {\tilde H}^{a5},\ {\tilde
  H}^{45}$, sourced by two scalar fields.  To describe bound states of
string solitons, operationally, we wrote down solutions to the BPS
equations that describe a superposition of two non-parallel string
solitons\ (the 6-dim theory has only self-dual strings so
``electric/magnetic'' simply label the 4/5-directions that the strings
are stretched along; the non-parallelness of the strings leads to
electric/magnetic in 4-dimensions upon 45-compactification). In the
simplest case of two orthogonal strings, these are
(\ref{m5m2-111001}), with behaviour as described subsequently. We then
impose boundary conditions describing the scalar moduli values as we
approach the string locations: these are imposed at the closest point
of approach of the two strings, given by (\ref{bc-111001}) in the
simplest case (analogous to (\ref{bc-M5spike}) for a single string but
with more features). This is equivalent to fixing the separation
between the strings transverse to their extended directions in terms
of the distance from the wall of marginal stability, as in
(\ref{111001-M5wms}). As we have seen, imposing these boundary
conditions or more generally (\ref{bc-gen}) at the closest approach
becomes an increasingly better approximation to imposing them all
along the strings (\ie\ at $\vrho_{\zeta^1}\ra\vrho_{\zeta^1}^0$ and
$\vrho_{\zeta^2}\ra\vrho_{\zeta^2}^0$), in the limit of approaching
the wall of marginal stability in moduli space, as we discussed in
detail around (\ref{bc-111001}). In this limit, the separation between
the solitons increases and eventually diverges
(\ref{111001-WMSlimit}), so the state then becomes unbound (and
disappears from the spectrum as one crosses the wall). While there are
clear parallels with string web states in super Yang-Mills theories,
it may be of importance to understand wall-crossing phenomena more
systematically for such string-string bound state molecules and the
role played by the extended nature of the constituents, for instance
in regard to their degeneracies which are likely to have more features
compared with \cite{Sen:2007ri,Dabholkar:2008tm}\ (more generally see
\eg\ \cite{Sen:2007qy,Dabholkar:2012zz} for discussions on
wall-crossing in relation to black hole microstate counting).

Our description of these bound states has been indirect in a sense: we
have used the BPS bound equations and imposed appropriate boundary
conditions whose intuition stems from the M5-M2-M2 brane geometry we
expect in the space transverse to the M5-brane (and corresponding
descriptions of string web states via D3-brane deformations
\cite{Argyres:2001pv,Narayan:2007tx}). Our equilibrium configurations
(reliable in the limit of approaching the wall of marginal stability)
suggest that the scalar forces balance the 3-form forces between the
constituent string solitons.  It would be most interesting to
understand the nature of these bound states more intrinsically in the
M5-brane tensor theory, \eg\ in terms of the way the 3-form field
lines behave for such composite configurations and the corresponding
forces, more directly, between the constituent string solitons.

A related point is the following. As we have seen, we have imposed
boundary conditions (\ref{bc-111001}) (and more generally
(\ref{bc-gen})) at the point of closest approach of the two strings:
this leads to the moduli variations and associated brane-bendings, as
in Table~\ref{tableXYbndy21}, which become refined as we approach
marginal stability. One might ask if, instead, one could impose
boundary conditions that amount to holding the strings fixed far along
them (at infinity), and allowing the rest of the two-string geometry
(at finite distances) to be determined by energetics. It would seem
that in this case the scalar configurations will be different from
(\ref{m5m2-111001}), leading to brane-bendings that reflect in the two
strings deformed towards each other in their near regions. This might
possibly be a different energy extremization than
(\ref{completiongen}) with different boundary terms held fixed (see
sec.~\ref{sec:Bog1str-os} for some comments on strict Bogomolny
completions versus the extra terms in (\ref{completiongen})).  In a
sense this might be more natural if we consider the dynamical problem
of a self-dual string probe moving in the background field of another
non-parallel self-dual string: the forces will be less severe far
along the string probe (at infinity) while the string bulk will deform
more.  In the limit of approaching marginal stability, from our
discussions of the field configurations, moduli variations and the
tension in sec.~\ref{sec:tension} and sec.~\ref{sec:Bog1str-os}, it is
clear that the above description will approach our formulation.  In
some sense this sort of string-string dynamics might be describable by
appropriate string uplifts of the particle dynamics in
\cite{Lee:2011ph}, \cite{Kim:2011sc}.  It would be instructive to
understand these issues better, and we hope to do so.

Perhaps understanding these string-string effective potentials could
be explored building on \cite{Denef:2002ru} (see
\eg\ \cite{Gustavsson:2001wa,Gustavsson:2008dy} for some interesting
related discussions). One might hope thereby to gain valuable insight
into the dynamics of self-dual string solitons and thence that of
self-dual 3-forms in the nonabelian M5-brane theory (see \eg\
\cite{Saemann:2017rjm,Rist:2022hci} for nonabelian self-dual strings,
which may bear on our discussions in sec.~\ref{M5s-higgsed} on
embedding into multiple abelian M5-brane theories). Relatedly, it
would be interesting to understand more systematically the role of
supersymmetry with regard to our Bogomolny completion, appropriately
generalizing \cite{Witten:1978mh}: this may dovetail with the role of
fermionic zero modes in the effective supersymmetric sigma model
for small string-string fluctuations.

It is worth noting that our Bogomolny completion is in a sense tied to
the existence of two $\zeta$-vectors in the $(x^4,x^5)$-plane: this
means the two M2-branes ending on the M5-brane stretch in this 2-dim
subspace alone so the 3-dim $\vr$-subspace is untouched. In some
essential sense, this dovetails with the intuition that the these
M5-M2-M2 configurations will compactify to become appropriate D3-brane
string-web configurations with the $\vr$-subspace being the D3-brane
worldvolume. On the other hand, one could imagine more general
configurations such as with three or more M2-branes ending on the
M5-brane (see \cite{Gauntlett:1999aw} for some investigations on
these). These may relate to the D3-brane prong configurations in
\cite{Argyres:2001pv,Narayan:2007tx} pertaining to three or more
charge centers (see \eg\ \cite{Denef:2007vg} for multiple center black
hole configurations). It would be interesting to explore these
further.

There are also a few other interesting questions that come to
mind. One pertains to generalizing these M5-brane prong configurations
to less supersymmetric theories such as those in
\eg\ \cite{Witten:1997sc} and \cite{Gaiotto:2009we}, possibly
interlinking with \cite{Gaiotto:2009hg}, \cite{Cecotti:2009uf}: these
might be related to D3-brane prong descriptions \cite{Argyres:2001pv}
of string webs \cite{Gaberdiel:1998mv,Bergman:1998br,Mikhailov:1998bx}
in \Nt\ SYM theories \cite{Seiberg:1994rs}, and might exhibit
appropriate s-rules in regard to the BPS spectrum.  Another pertains
to whether these sorts of string soliton bound states in the 6-dim
$(2,0)$ theory can be realized explicitly as field configurations in
the nonperturbative 5-dim SYM theory description discussed in
\cite{Douglas:2010iu,Lambert:2010iw}.  A third involves understanding
5-brane webs \eg\ \cite{Aharony:1997bh} (see also
\cite{Bergman:2015dpa}) as connected prong-like brane-deformations. A
still further question has to do with whether an M2-brane web
(regarded as an object in itself, with no M5-brane) can be described
from the point of view of the M2-brane worldvolume theory itself
\cite{Bagger:2007jr,Gustavsson:2007vu,Aharony:2008ug,Aharony:2008gk}:
in some sense this might be analogous to the description of a string
web (in the absence of a D3-brane) from the point of view of the
D-string worldvolume gauge theory \cite{Dasgupta:1997pu}. Perhaps more
general M2-brane networks in M-theory might be related to IIB string
networks \cite{Sen:1997xi}.

Our analysis of the M5-brane prong configurations may also be useful
in understanding other BPS observables in the 6d worldvolume
theory. The analysis of surface observables in
\cite{Drukkeretal:0420}, \cite{FGT:2015} uses a similar M2-M5 brane
construction to study various geometric features and non-perturbative
aspects associated with them in the 6d $(2,0)$ theory. In this regard,
our single M5-brane spike configuration is similar to the half-BPS
surface observable in these references. But while those works involved
the use of $AdS_7 \times S^4$ holography \cite{Drukkeretal:0420} or
used geometric constructions on special manifolds \cite{FGT:2015}, a
detailed analysis in the theory on a single M5 worldvolume is not done
in this way. It would be interesting to see if our configurations with
the M5-prong structures help uncover more about these BPS surface
observables.

\vspace{6mm}

{\footnotesize \noindent {\bf Acknowledgements:}\ \ It is a pleasure
  to thank Philip Argyres, Sujay Ashok, Jerome Gauntlett and Costis
  Papageorgakis for helpful comments on a draft.  This work is
  partially supported by a grant to CMI from the Infosys Foundation.}

\vspace{-2mm}

\appendix

\section{Useful identities}\label{AppA:identities}

This appendix contains some details on intermediate steps involved in
obtaining the Bogomolnyi completion in (\ref{completiongen}) from the
energy functional expression (\ref{freelimEngy}).
	\begin{itemize}
		\item First we have
		\begin{align}
	 \left| \partial_{[a} X \zeta^{(1)}_{b]} \right|^2 = & \sum_{a \ne b \,;\, c \ne d} \frac{\delta^{ac} \delta^{bd}}4 \left( \partial_a X \zeta^{(1)}_b - \partial_b X \zeta^{(1)}_a\right) \left( \partial_c X \zeta^{(1)}_d - \partial_d X \zeta^{(1)}_c\right) \cr
				=	& \,\, \frac 12 \sum_{a \ne b }  \left( \partial_a X \right)^2 \left( \zeta^{(1)}_b \right)^2 \,-\, \frac 12 \sum_{a \ne b }  \left( \partial_a X \zeta^{(1)}_a \right) \left( \partial_b X \zeta^{(1)}_b \right)
		\end{align}
\item The expansion of $\left( \partial_a X \zeta^{(1)}_a \right)^2$ is the following
		\begin{align}
			 	\sum_{a,b} \left( \partial_a X \zeta^{(1)}_a \right) \left( \partial_b X \zeta^{(1)}_b \right) 
			 =  \sum_{a }  \left( \partial_a X \right)^2 \left( \zeta^{(1)}_a \right)^2 \,+\,  \sum_{a \ne b }  \left( \partial_a X \zeta^{(1)}_a \right) \left( \partial_b X \zeta^{(1)}_b \right)
		\end{align}
	\end{itemize}	
	And the sum $\left| \partial_{[a} X \zeta^{(1)}_{b]} \right|^2 \,+\, \frac 12 \times \left( \partial_a X \zeta^{(1)}_a \right)^2$ is equal to $\frac 12 \left| \partial X \right|^2 $.
	\begin{itemize}
		\item The dot product $( \partial_{[a} X \zeta^{(1)}_{b]} ) \cdot ( \partial_{[a} Y \zeta^{(2)}_{b]} )$ can be expanded as
		\bea
		&&	 \ \partial_{[a} X \zeta^{(1)}_{b]} \partial_{[c} Y \zeta^{(2)}_{d]} \delta^{ac} \delta^{bd}
			 =  \sum_{a \ne b \,;\, c \ne d} \frac{\delta^{ac} \delta^{bd}}4 \left( \partial_a X \zeta^{(1)}_b - \partial_b X \zeta^{(1)}_a\right) \left( \partial_c Y \zeta^{(2)}_d - \partial_d Y \zeta^{(2)}_c\right) \nn\\
		&&	\qquad\qquad\quad = \,\, \frac 12 \sum_{a \ne b }  \left( \partial_a X \partial_a Y  \right)  \left(  \zeta^{(1)}_b \zeta^{(2)}_b \right) \,-\, \frac 12 \sum_{a \ne b }  \left( \partial_a X \zeta^{(2)}_a \right) \left( \partial_b Y \zeta^{(1)}_b \right)
		\eea
		\item And the expansion of $\left( \partial_a X \zeta^{(1)}_a \right) \left( \partial_b Y \zeta^{(2)}_b \right)$ is the following
		\be 
			\sum_{a,b} \left( \partial_a X \zeta^{(1)}_a \right) \left( \partial_b Y \zeta^{(2)}_b \right) 
		=	 \,\,  \sum_{a }  \left( \partial_a X \partial_a Y  \right)  \left(  \zeta^{(1)}_a \zeta^{(2)}_a \right) \,+\, \sum_{a \ne b }  \left( \partial_a X \zeta^{(1)}_a \right) \left( \partial_b Y \zeta^{(2)}_b \right) \,.
		\ee 
	\end{itemize}
	So the addition: $2 \times ( \partial_{[a} X \zeta^{(1)}_{b]} ) \cdot ( \partial_{[a} Y \zeta^{(2)}_{b]} ) \,+ \, \left( \partial_a X \zeta^{(1)}_a \right) \left( \partial_b Y \zeta^{(2)}_b \right)$ gives the following answer
	\begin{align}
			\sum_{a ,b }  \left( \partial_a X \partial_a Y  \right)  \left(  \zeta^{(1)}_b \zeta^{(2)}_b \right)  \, + \, \sum_{a\ne b} \zeta^{(1)}_a \zeta^{(2)}_b \left( \partial_a X \partial_b Y - \partial_b X \partial_a Y\right) \,.
	\end{align}

\section{A brief review: D3-brane prongs and string webs}\label{sec:D3prongs}

Here we give a short recap of the D3-brane prong description of string
web states in \cite{Argyres:2001pv,Argyres:2000xs,Narayan:2007tx}.
The low energy abelian effective action on a D3-brane leads to an energy
functional\ $\vE^2+\vB^2+(\nabla X)^2+(\nabla Y)^2$\,: integrating over
3-space with charge boundary conditions gives the mass for the
configuration as
\be\label{Bogomolny:D3}
{\cal M}
= \ \int d^3x\,\frac12 \left[(\vE-\nabla X')^2+(\vB-\nabla Y')^2\right] 
+ \oint_{S^2_I} d{\vec a}\cdot \big(\vE\, X' + \vB\, Y'\big)\, 
\ee
where $X'=X\cos\varphi+Y\sin\varphi$ and $Y'=-X\sin\varphi+Y\cos\varphi$.
The boundaries $S^2_I$ are 2-spheres around each of several charge 
cores in the system and one sphere at infinity, and we have used the 
divergence-free equations for the electric and magnetic fields 
$\vn\cdot\vE=\vn\cdot\vB=0$ away from the cores to write the boundary
term. We label the $I=0$ boundary as the one at infinity and the
$I\neq 0$ ones as those around the charge cores. Within the abelian
approximation, the scalars $X, Y$ can be regarded as taking constant
values $X_i, Y_i$ at the $i$-th boundary while at infinity they take
their asymptotic vacuum values $(-X^0,-Y^0)$. With the electric/magnetic
charges\ $\oint_{S^2_i}\vE\cdot d{\vec a}=Q_E^i\ ,
\ \oint_{S^2_i}\vB\cdot d{\vec a}=Q_B^i$, 
of the various cores, the charges at infinity are\
$(Q_E^0,Q_B^0) = -\sum_{i=1}^n (Q_E^i,Q_B^i)$,\ by charge conservation.
Then (\ref{Bogomolny:D3}) becomes
\be
\sum_i\ \cos\varphi \big( (X_i+X_0)Q_E^i + (Y_i+Y_0)Q_B^i \big)\,
+\, \sin\varphi \big( (Y_i+Y_0)Q_E^i - (X_i+X_0)Q_B^i \big)\ .
\ee
Extremizing  with the angle $\varphi$ gives the tightest bound 
(with a condition analogous to (\ref{varphiXY})), thereby leading
to the BPS mass formula
\be
{\cal M} = \sqrt{ \big( (X_i+X_0)Q_E^i + (Y_i+Y_0)Q_B^i \big)^2
  + \big( (Y_i+Y_0)Q_E^i - (X_i+X_0)Q_B^i \big)^2 }\,.
\ee
This is essentially identical to the right hand side of (\ref{T-bound})\
(the scalars above have mass dimension one though, unlike in the
M5-case; see comments after (\ref{M5-XI-fieldthLim})).
Although we have an effective particle bound state mass here, rather
than the effective string tension in (\ref{T-bound}), the algebra of
the Bogomolny bound, in particular (\ref{extr-varphi}),
(\ref{varphiXY}), leading to ${\cal M}$ is similar (for more
details, see \cite{Argyres:2001pv,Argyres:2000xs,Narayan:2007tx}, and
earlier work \eg\ \cite{Bergman:1997yw,Bergman:1998gs,Lee:1998nv}).\
As an example, a single core dyon charge $(Q_E,Q_B)$ and a single
scalar $X$ gives\ ${\cal M} = \sqrt{Q_E^2+Q_B^2}\,(X_1+X_0)$.
For the simple case with one electric and one magnetic particle, we
have $(Q_E^1,Q_B^1)=(q_e,0)$ and $(Q_E^2,Q_B^2)=(0,q_m)$, with moduli
values\ $(X_1,0)$ and $(0,Y_1)$. So this gives for $\varphi$ an
expression identical to (\ref{phi-0:modConst}), amounting to an
overall rotation of $X,Y$, which leads to an analog of the moduli
constraint \eg\ (\ref{moduliConstr-111001}) mapping the charge to
the asymptotic shape of the D3-brane deformation.
We thus obtain the BPS bound equations
\be
\vE=\nabla X\ ,\qquad \vB=\nabla Y\ ;\qquad
\nabla\cdot\vE=\nabla^2X=0\ ,\quad \nabla\cdot\vB=\nabla^2Y=0\ .
\ee
so the scalars $X, Y$ are harmonic in the D3-brane spatial worldvolume.
The rotation to $\phi=0$ implies that electric charge configurations
are sourced by nontrivial $X$ scalar deformations of the D3-brane
worldvolume (in the transverse $X$-direction) while magnetic
ones are sourced by nontrivial $Y$ scalar D3-brane deformations. Generic
solutions to these represent string web states in the gauge theory.\
A simple example is the field configuration 
\be
X = {q_e\over |\vr-\vr_1|} - X_0\,,\qquad Y = {q_m\over |\vr-\vr_2|} - Y_0\,,
\ee
representing a charge $(1,1)-(1,0)-(0,1)$ state, with one electric
and one magnetic charge center. Its asymptotic behaviour is
\be
|\vr|\ra\infty:\qquad X\sim {q_e\over|\vr|}-X_0\,,\quad
Y\sim {q_m\over|\vr|}-Y_0\,,
\ee
so $(\vE,\vB)\propto (1,1)$ encodes charge $(1,1)$ in $q_e,q_m$ units.
This deformation traces a line along $(1,1)$ in the moduli space if
${1\over q_e}X_0={1\over q_m}Y_0$.\ Further as we approach either
charge core, we see 
\be
\vr\ra\vr_1:\quad Y\sim {q_m\over |\vr_1-\vr_2|}-Y_0\,;\qquad\quad
\vr\ra\vr_2:\quad X\sim {q_e\over |\vr_1-\vr_2|}-X_0\,,
\ee
while $X\ra X_1$ and $Y\ra Y_1$ develop long spikes respectively
(which we have cut off). It is then consistent to impose the
boundary conditions on the moduli as
\be
\vr\ra\vr_1:\quad (X,Y)\ra (X_1,0)\,;\qquad\quad
\vr\ra\vr_2:\quad (X,Y)\ra (0,Y_1)\,.
\ee
This leads to fixing the separation between the charge cores
\be
{1\over|\vr_1-\vr_2|}\sim {1\over q_m}Y_0 = {1\over q_e}X_0\ ,
\ee
which is large relative to the core sizes (which scale as
${1\over X_1}$ and ${1\over Y_1}$) in the limit of approaching the
wall of marginal stability, with $X_0, Y_0\ \ll\ X_1, Y_1$.\
As we hit this wall, \ie\ $X_0, Y_0\ra 0$, the state becomes
arbitrarily loosely bound and decays. No bound state solutions exist
on the other side of the wall (for $X_0<0$).

More general field configurations can be written as solutions to the
BPS bound equations, with similar reliable behaviour near the wall
of marginal stability.

More details on mapping these low energy field theory prong
configurations to string webs stretched in the transverse space
between D-branes appears in
\cite{Argyres:2001pv,Argyres:2000xs,Narayan:2007tx}.

\section{Some details on more general M5-M2 states}\label{App:details2132}

In this appendix we give some details of use in analyzing the brane
deformations presented in the main section (\ref{sec:m5m2gen}). The
expression for the ``basis fields'' $X$ and $Y$ are the following
\be\label{m5m2-XY}
X = {c_X\over (\vr-\vr_1)^2 + (\zeta_1^\perp\cdot  \vx )^2} - X_0\ ,\qquad \quad
Y = {c_Y\over (\vr-\vr_2)^2 + (\zeta_2^\perp\cdot \vx )^2} - Y_0 \,  \hspace{2cm}
\ee

\subsection{(2,1) - (1,0) - (1,1)}

For the $(2,1) - (1,0) - (1,1)$ M5-prongs we consider $X$, $Y$ with $c_X = q_e$ and $c_Y = q_m / \gamma$. 

\noindent
At large radial distance from the string sources $|\vec{r}| \rightarrow \infty$ these ``basis'' fields approach
\be
|\vec{r}| \rightarrow \infty\,:
\qquad X \sim - X_0 \quad \qquad Y \sim - Y_0 \,.
\ee
As we approach $\text{string}_1$, with coordinates in
(\ref{211011-zetaStringCoordsa}), the $X$ field develops a spike while the
$Y$ field is nonsingular:
\be
\vec{\rho} \rightarrow \left( \vec{r}_1, x_1, 0 \right)\,: \qquad X \rightarrow X_1 \qquad \quad Y \sim \frac{c_Y}{\left( \vec{r}_1 - \vec{r}_2\right)^2 + \gamma^2 \left( x_1 \right)^2} - Y_0
\ee
Analogous to the $(1,1) - (1,0) - (0,1)$ example, at $(r_1, 0, 0)$ we impose
\be
\label{condition21Xspike}
Y = 0 \quad \implies  \quad \frac{c_Y}{\left( \vec{r}_1 - \vec{r}_2\right)^2} = Y_0
\ee
On approaching $\text{string}_2$ (onto (\ref{211011-zetaStringCoordsb}))
the $Y$ field develops a spike while $X$ is nonsingular:
\be
\vec{\rho} \rightarrow \left( r_2 \, , \beta  \, x_2 \,, \, x_2 \right)\,: \qquad X \sim {q_e\over (\vr_2-\vr_1)^2 + (x_2)^2} - X_0 \qquad \quad Y \rightarrow Y_1\ .
\ee
So at $(r_2, 0, 0)$, we impose
\be
\label{condition21Yspike}
X = 0 \quad \implies  \quad \frac{q_e}{\left( \vec{r}_2 - \vec{r}_1\right)^2} = X_0
\ee
\bc
\begin{figure}[H]
	\bc\includegraphics[width=20pc]{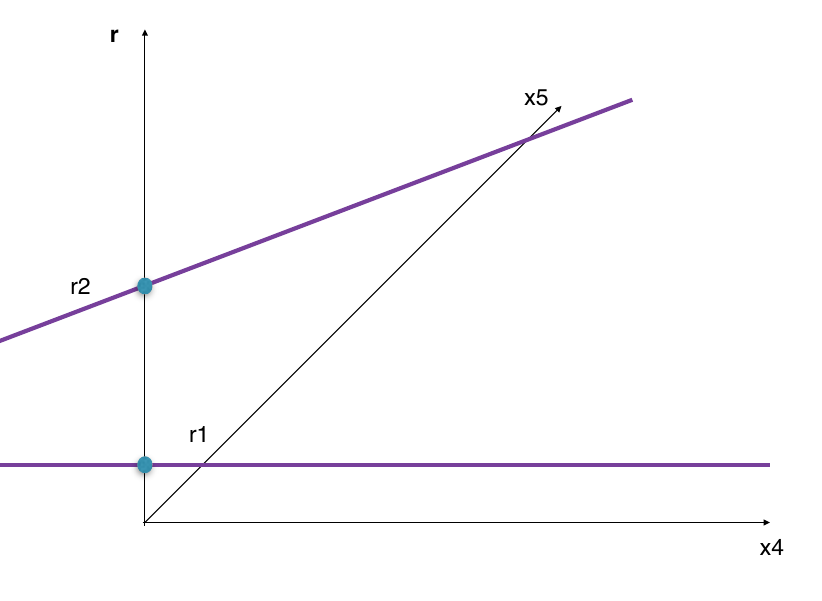}\ec
	\caption{{ \label{twolinesources}
			\footnotesize This depicts the orientations and the positions of the two strings in the 5-dimensional space on the worldvolume. The two strings are not orthogonal when projected on the $x^4$-$x^5$ plane.
	}}
\end{figure}
\ec
Here we see that the values of $X$ and $Y$ fields interpolate as we move along the two strings away from the origin. These values are tabulated in
Table~\ref{tableXYbndy21}, except they are to be interpreted as values
for the ``basis'' fields for our purposes now.\
Using these asymptotics for $X, Y$ and the relations (\ref{XYvsXpYp21}),
we obtain the asymptotics for $X', Y'$ in Table~\ref{tableXYprimebndy}.

\subsection{(3,2) - (1,1) - (2,1)}

For the $(3,2) - (1,1) - (2,1)$ prongs we consider $X$, $Y$ in (\ref{m5m2-XY}) with $c_X = q_e / \gamma_1$ and $c_Y = q_m / \gamma_2$. \\

\medskip
\noindent
When the radial distance is large from the string sources $|\vec{r}| \rightarrow \infty$ these ``basis'' fields approach $(-X_0,-Y_0)$ as before.
As in the previous example as we approach $\text{string}_1$ with
coordinates in (\ref{321121-zetaStringCoordsa})), the $X$ field develops
a spike and the $Y$ field is nonsingular:
\be
\vec{\rho} \rightarrow \left( \vec{r}_1, x_1, \beta_1  x_1 \right)\,: \qquad X \rightarrow X_1 \qquad \quad Y \sim \frac{c_Y}{\left( \vec{r}_1 - \vec{r}_2\right)^2 + \gamma_2^2 \left( x_1 \right)^2} - Y_0
\ee
%
Analogous to the previous example, at $(r_1, 0, 0)$ we impose
\be
\label{condition32Xspike}
Y = 0 \quad \implies  \quad \frac{c_Y}{\left( \vec{r}_1 - \vec{r}_2\right)^2} = Y_0
\ee
On approaching $\text{string}_2$ (see (\ref{321121-zetaStringCoordsb}))
there is a spike in the value of the $Y$ field while the $X$ field is
nonsingular:
\be
\vec{\rho} \rightarrow \left( r_2 \, , \beta_2  \, x_2 \,, \, x_2 \right)\,: \qquad X \sim {c_X\over (\vr_2-\vr_1)^2 + \gamma_1^2 (x_2)^2} - X_0 \qquad \quad Y \rightarrow Y_1
\ee
So at $(r_2, 0, 0)$, we impose
\be
\label{condition32Yspike}
X = 0 \quad \implies  \quad \frac{c_X}{\left( \vec{r}_2 - \vec{r}_1\right)^2} = X_0
\ee
Again we see that the values of $X$ and $Y$ fields interpolate as we
move along the two strings away from the origin.  Although the
detailed values of $X, Y$ change in the regions along the strings, in
the asymptotic regions they become the same as in
Table~\ref{tableXYbndy21}.\
Using these and the relations (\ref{XYvsXpYp32}), we obtain the
asymptotics for $X', Y'$ in Table~\ref{tableXYprimebndy32}.

\subsection{1-string tension}\label{sec:App-tension2132}

First we consider the $(2,1) - (1,0) - (1,1)$ web. The boundary terms in
(\ref{BogomolnyCompnnts}) upon integrating become
\be
\int dx^4 dx^5	\int d^3r  \,\, \tilde{\mathcal{H}}^{a4}  \partial_a X'  \, + \,\int dx^4 dx^5	\int d^3r   \,\, \tilde{\mathcal{H}}^{a5}  \, \partial_a Y'
\ee
In terms of  $X$, $Y$ fields this is
\begin{align}\label{noncrossingbndytrm(2,1)}
\int dx^4 dx^5	\int d^3r  \,\, \tilde{\mathcal{H}}^{a4} \left(  \partial_a X \,+\,  \beta \gamma \, \partial_a Y  \right)  \, + \,\int dx^4 dx^5	\int d^3r   \,\, \tilde{\mathcal{H}}^{a5}  \,\gamma \, \partial_a Y
\end{align}
For convenience we write down the explicit expression for these fields
\be\label{m5m2-XY'}
X = {q_e\over (\vr-\vr_1)^2 + (x^5 )^2} - X_0\ ,\qquad \quad
Y = {q_m / \gamma \over (\vr-\vr_2)^2 + \gamma^2 ( x^4 - \beta x^5 )^2} - Y_0 \,  \hspace{2cm}
\ee
In (\ref{noncrossingbndytrm(2,1)}) we do the integration by parts and
subsequently use the equation of motion\ 
 $\partial_a \tilde{\mathcal{H}}^{ab} = 0$\ to get
\begin{align}
\int dx^4 \oint_{S^3} d\hat{s}^a \,\, \tilde{\mathcal{H}}^{a4}  \,  X \, + \, \beta \gamma \int d\tilde{x}^5	\oint_{S^3} d\hat{s}^a \,\, \tilde{\mathcal{H}}^{a4}  \,  Y  \, + \, \gamma \int d\tilde{x}^5	\oint_{S^3} d\hat{s}^a \,\, \tilde{\mathcal{H}}^{a5}  \,  Y
\end{align}
where $d\tilde{x}^5$ is the differential length element along string$_2$
and should be expressible as: \\
$d\tilde{x}^5 = \gamma \left(  \beta dx^4 \, + \, dx^5 \right)$.\ 
We are considering two strings with charges
\be
(Q_E^i,Q_B^i),\ i=1,2 :\qquad
Q_E^i=\oint_{S^3_{i}} {\tilde { \mathcal H}}^{a4} ds^a\ ,\qquad
Q_B^i=\oint_{S^3_{i}} {\tilde { \mathcal H}}^{a5} ds^a\ ,
\ee
and scalar moduli values at their cores as given in
Table (\ref{tableXYbndy21}). At spatial infinity far from both strings,
charge conservation gives
$(Q_E^0,Q_B^0)=-\sum_i (Q_E^i,Q_B^i)$ and the moduli values are
$(-X_0,-Y_0)$.
\begin{align}
& \int dx^4 \oint_{S^3_{i=1}} d\hat{s}^a \,\, \tilde{\mathcal{H}}^{a4}  \,  X \, + \, \beta \gamma \int d\tilde{x}^5	\oint_{S^3_{i=2}} d\hat{s}^a \,\, \tilde{\mathcal{H}}^{a4}  \,  Y  \, + \, \gamma \int d\tilde{x}^5	\oint_{S^3_{i=2}} d\hat{s}^a \,\, \tilde{\mathcal{H}}^{a5}  \,  Y \cr 
&\ + \, \int dx^4	\oint_{S^3_{\infty}} d\hat{s}^a \,\, \tilde{\mathcal{H}}^{a4}  \,  X \, + \, \beta \gamma \int d\tilde{x}^5	\oint_{\tilde{S}^3_{\infty}} d\hat{s}^a \,\, \tilde{\mathcal{H}}^{a4}  \,  Y  \, + \, \gamma \int d\tilde{x}^5	\oint_{\tilde{S}^3_{\infty}} d\hat{s}^a \,\, \tilde{\mathcal{H}}^{a5}  \,  Y \cr
\end{align}
Integrating over 5-space, and so at appropriate $S^3_{i}$, gives
\be
L \left( \left(X_1+X_0 + \beta \gamma \, Y_0\right)Q_E^1 \,+\, \beta \gamma \left(Y_1+Y_0 +{ X_0 \over  \beta \gamma }\right) Q_E^2 \,
+\,  \gamma (Y_1+Y_0)Q_B^2  \right)\ ,
\ee
with $L = \int dx^4 \sim \int d\tilde{x}^5$ the regulated length of both
strings.\ \ 
Since $\left(Q^1_E, Q^1_B \right) = ( q_e , 0 )$ and 
$\left(Q^2_E, Q^2_B \right) = ( q_e , q_m )$ this is equal to 
\be
L \left( \left(X_1+X_0 + \beta \gamma \, Y_0\right)q_e \,+\, \beta \gamma \left(Y_1+Y_0 +{ X_0 \over  \beta \gamma }\right) q_e \,
+\,  \gamma (Y_1+Y_0)q_m  \right)\ 
\ee
Further, the above formula can also be written in terms of $X'$ and $Y'$ field values as given in (\ref{mass-tension(2,1)}) in section \ref{tensionnon-ortho}.

\bigskip

Likewise, for the $(3,2) - (1,1) - (2,1)$ web, the boundary terms upon
integrating become
\begin{align}
\int dx^4 dx^5	\int d^3r  \,\, \tilde{\mathcal{H}}^{a4}  \partial_a X'  \, + \,\int dx^4 dx^5	\int d^3r   \,\, \tilde{\mathcal{H}}^{a5}  \, \partial_a Y'
\end{align}
In terms of  $X$, $Y$ fields this is
\begin{align}\label{noncrossingbndytrm(3,2)}
\int dx^4 dx^5	\int d^3r  \,\, \tilde{\mathcal{H}}^{a4} \left(  \gamma_1 \partial_a X \,+\,  \beta_2 \gamma_2 \, \partial_a Y  \right)  \, + \,\int dx^4 dx^5	\int d^3r   \,\, \tilde{\mathcal{H}}^{a5} \left( \beta_1 \gamma_1 \, \partial_a X \, + \,\gamma_2 \, \partial_a Y \right)
\end{align}
For convenience we write down the explicit expression for these fields
\be\label{m5m2-XY(3,2)}
X = {q_e / \gamma_1 \over (\vr-\vr_1)^2 + \gamma_1 ( \beta_1 x^4 - x^5 )^2} - X_0\ ,\qquad \quad
Y = {q_m  / \gamma_2 \over (\vr-\vr_2)^2 + \gamma_2^2 ( x^4 - \beta_2 x^5 )^2} - Y_0 \,  \hspace{2cm}
\ee
In (\ref{noncrossingbndytrm(2,1)}) we do the integration by parts and
subsequently use the equation of motion
$\partial_a \tilde{\mathcal{H}}^{ab} = 0$ to get
\begin{align}
& \gamma_1 \int d\tilde{x}^4	\oint_{S^3} d\hat{s}^a \,\, \tilde{\mathcal{H}}^{a4}  \,  X \, + \, \beta_2 \gamma_2 \int d\tilde{x}^5	\oint_{S^3} d\hat{s}^a \,\, \tilde{\mathcal{H}}^{a4}  \,  Y \, + \, \beta_1 \gamma_1 \int d\tilde{x}^4	\oint_{S^3} d\hat{s}^a \,\, \tilde{\mathcal{H}}^{a5}  \,  X \cr
& + \, \gamma_2 \int d\tilde{x}^5	\oint_{S^3} d\hat{s}^a \,\, \tilde{\mathcal{H}}^{a5}  \,  Y
\end{align}
where $d\tilde{x}^4$ and $d\tilde{x}^5$ are the differential length
elements along string$_1$ and string$_2$ respectively, and should be
expressible as: 
\be
d\tilde{x}^4 = \gamma_1 \left(   dx^4 \, + \, \beta_1 \, dx^5 \right) ,\qquad
d\tilde{x}^5 = \gamma_2 \left(  \beta_2 \, dx^4 \, + \, dx^5 \right) . \nn
\ee
As before, we are considering two string charges
$(Q_E^i,Q_B^i),\ i=1,2$, with scalar moduli values at their cores as
given in Table (\ref{tableXYbndy21}). At spatial infinity far from both
strings, charge conservation gives $(Q_E^0,Q_B^0)=-\sum_i (Q_E^i,Q_B^i)$.
We obtain
\begin{align}
\, & \textstyle \gamma_1 \int d\tilde{x}^4	\oint_{S^3_{i=1}} d\hat{s}^a \,\, \tilde{\mathcal{H}}^{a4}  \,  X \, + \, \beta_1 \gamma_1  \int d\tilde{x}^4 \oint_{S^3_{i=1}} d\hat{s}^a \,\, \tilde{\mathcal{H}}^{a5}  \,  X \, + \, \beta_2 \gamma_2 \int d\tilde{x}^5	\oint_{S^3_{i=2}} d\hat{s}^a \,\, \tilde{\mathcal{H}}^{a4}  \,  Y  \, \cr 
& \textstyle + \, \gamma_2 \int d\tilde{x}^5	\oint_{S^3_{i=2}} d\hat{s}^a \,\, \tilde{\mathcal{H}}^{a5}  \,  Y 
\, + \,  \gamma_1 \int d\tilde{x}^4	\oint_{S^3_{\infty}} d\hat{s}^a \,\, \tilde{\mathcal{H}}^{a4}  \,  X \, + \, \beta_1 \gamma_1  \int d\tilde{x}^4 \oint_{S^3_{\infty}} d\hat{s}^a \,\, \tilde{\mathcal{H}}^{a5}  \,  X \cr 
& \textstyle \, + \, \beta_2 \gamma_2 \int d\tilde{x}^5	\oint_{\tilde{S}^3_{\infty}} d\hat{s}^a \,\, \tilde{\mathcal{H}}^{a4}  \,  Y  \, 
+ \, \gamma_2 \int d\tilde{x}^5	\oint_{\tilde{S}^3_{\infty}} d\hat{s}^a \,\, \tilde{\mathcal{H}}^{a5}  \,  Y 
\end{align}
Integrating over 5-space, and so at appropriate $S^3_{i}$, gives
\begin{align}
& L \left[ \left(\gamma_1 X_1+ \gamma_1 X_0 + \beta_2 \gamma_2 \, Y_0\right)Q_E^1 \, + \, \left( \beta_1 \gamma_1 \, X_1 + \beta_1 \gamma_1 \, X_0 + \gamma_2 \, Y_0  \right) Q_B^1 \right] \cr 
 \,& +\, L \left[ \left( \beta_2 \gamma_2 \, Y_1+ \beta_2 \gamma_2 \, Y_0 + \gamma_1 X_0 \right) Q_E^2 \,
+\,   (\gamma_2 \, Y_1+ \gamma_2 \, Y_0 + \beta_1 \gamma_1 \, X_0 )Q_B^2  \right]\ ,
\end{align}
with $L = \int d\tilde{x}^4 \sim \int d\tilde{x}^5$ the regulated length of both
strings.

\noindent
The charge associated here are: $\left(Q^1_E, Q^1_B \right) = ( q_e , q_m )$; $\left(Q^2_E, Q^2_B \right) = (2  q_e , q_m )$; and we re-write the above formula using the $X'$ and $Y'$ field values to obtain (\ref{mass-tension(3,2)}) in section \ref{tensionnon-ortho}.

\bigskip

\noindent
{\large\textbf{Cross term contribution}}

There is another non-trivial contribution to the tension formula from
terms in (\ref{BogomolnyCompnnts}). The term $\left(\zeta^1 \cdot \zeta^2\right) \left( \partial X \cdot \partial Y \right)$ does not vanish for the configurations for which the two string sources are non-orthogonal. Under the 5-space integral we do the integration by parts so
\be
{\cal T}_{cr} \, = \, \int d \hat{x}_i  \oint_{S^3_i} ds^a  \left( \zeta^1 \cdot \zeta^2 \right) \, \partial_a X \, Y  \,-\, \int d^5x \, \partial^2 X \, Y \,.   
\ee
Here $d\hat{x}_i$ is the differential length element associated with the $i^{th}$ string charge core and $S^3_i$ is the hypersphere that encloses it. The second integral above which is over bulk 5d space does not contribute as the value of $\partial^2 X$ is non-zero only at the locations of the charge cores where we have defined the boundary for our integration region. 

	
For the $(2,1) - (1,0) - (1,1)$ configuration we have two boundary regions which enclose the two string charge cores respectively, and the remaining boundaries are at the asymptotic regions near infinity. Taking account of this we can write ${\cal T}_{cr}$ as
\begin{align}
{\cal T}_{cr} \, = \, &  \gamma \beta \, \int dx^4  \oint_{S^3_{i=1}} ds^a  \, \partial_a X \, Y    \, + \, \gamma \beta \, \int d\tilde{x}^5  \oint_{S^3_{i=2}} ds^a  \, \partial_a X \, Y \cr 
&  + \,  \gamma \beta \, \int dx^4  \oint_{S^3_{\infty}} ds^a  \, \partial_a X \, Y    \, + \, \gamma \beta \, \int d\tilde{x}^5  \oint_{\tilde{S}^3_{\infty}} ds^a  \, \partial_a X \, Y
\end{align}
Given that the value of the $Y$ field on $\text{string}_1$ interpolate
from $0$ at $x^4 = 0$ to $-Y_0$ at $|x^4| \rightarrow \infty$ and the
value of the $X$ field on $\text{string}_2$ from $0$ at the centre to
$-X_0$ far along the string, we can approximate ${\cal T}_{cr}$ as
(for small $X_0$, $Y_0$ values)
\begin{align}
{\cal T}_{cr} \, \sim \,\, \, &  -  \gamma \beta \, \int dx^4  \oint_{S^3_{i=1}} ds^a  \, \partial_a X \, Y_0    \, + \, \gamma \beta \, \int d\tilde{x}^5  \oint_{S^3_{i=2}} ds^a  \, \partial_a X \, Y_1 \cr 
&  - \,  \gamma \beta \, \int dx^4  \oint_{S^3_{\infty}} ds^a \, \partial_a X \, Y_0    \, - \, \gamma \beta \, \int d\tilde{x}^5  \oint_{\tilde{S}^3_{\infty}} ds^a  \, \partial_a X \, Y_0
\end{align}
In the asymptotic space region the third and fourth contributions vanish
since the value of $X$ field is fairly constant and equal to its moduli
value $- X_0$ so $\partial X \sim 0$ in this region. Thus 
\begin{align}
{\cal T}_{cr} \, \sim \,\, \, & - \gamma \beta \, \int dx^4  \oint_{S^3_{i=1}} ds^a  \, \partial_a X \, Y_0    \, + \, \gamma \beta \, \int d\tilde{x}^5  \oint_{S^3_{i=2}} ds^a  \, \partial_a X \, Y_1 
\end{align}
The first term is vanishingly small near marginal stability $Y_0\sim 0$\
(and the $X$ terms involve a cutoff near string-1 and so are nonsingular).
For the second term, it is convenient to note that the $X$ field from
string$_1$ has its largest value at the closest point between string$_1$
and string$_2$, \ie\ at $x^4, x^5=0$\ (elsewhere $X$ is smaller, as is
$\del X$).\ So we approximate 
\be
\del X\Big|_{\text{string}_2} = \del \frac{q_e}{ \left( \vr - \vr_1 \right)^2 + \left( x^5\right)^2 }\Big|_{\text{string}_2}\ \sim\ \nabla_{\vr}\, \frac{q_e}{ \left( \vr - \vr_1 \right)^2 }\Big|_{\text{string}_2}
\ \sim\ -q_e {\vr_2-\vr_1\over \left( \vr_2 - \vr_1 \right)^4}
\ee
%
In the limit of large transverse separation $| \vr_1 - \vr_2 | $
between the two strings, \ie\  $X_0$, $Y_0$ $\ra 0$ near the wall of
marginal stability, this term is also vanishingly small. Thus
${\cal T}_{cr}$ tends to zero and there is no contribution
from this cross term to the mass formula in this limit.


\end{document}